\documentclass{article}

\usepackage{amssymb}
\usepackage{amsmath}
\usepackage{caption}
\usepackage{subfigure}
\usepackage{multicol}
\usepackage{here}
\usepackage{graphicx}
\usepackage{color}
\usepackage{algorithmicx}
\usepackage{algpseudocode}
\usepackage{algorithm}
\usepackage{listings}
\usepackage{epstopdf}
\usepackage{marvosym}
\usepackage{authblk}
\usepackage{mathtools}

\DeclareGraphicsExtensions{.eps}
\def\calN{\mathcal{N}}

\def\fatx{\mathbf{x}}

\def\fatx{\mathbf{x}}

\numberwithin{equation}{section}
\numberwithin{table}{section}
\numberwithin{figure}{section}

\begin{document}

\title{Excluded volume effects in on- and off-lattice reaction-diffusion models}
\author{Lina Meinecke\thanks{lina.meinecke@it.uu.se}, Markus Eriksson}
\affil{Department of Information Technology, Uppsala University}
\maketitle

\begin{abstract}
Mathematical models are important tools to study the excluded volume effects on reaction-diffusion systems, which are known to play an important role inside living cells.
Detailed microscopic simulations with off-lattice Brownian dynamics become computationally expensive in crowded environments.
In this paper we therefore investigate to which extent on-lattice approximations, so called Cellular Automata models, can be used to simulate reactions and diffusion in the presence of crowding molecules.
We show that the diffusion is most severely slowed down in the off-lattice model, since randomly distributed obstacles effectively exclude more volume than those ordered on an artificial grid.
Crowded reaction rates can be both increased and decreased by the grid structure and it proves important to model the molecules with realistic sizes when excluded volume is taken into account.
The grid artifacts increase with increasing crowder density and we conclude that the computationally more efficient on-lattice simulations are accurate approximations only for low crowder densities.
\end{abstract}

\section{Introduction}
Living cells are regulated by complicated signaling pathways that control which genes are expressed and how the cells behave.
These reaction networks are spatially organized with important reaction complexes often bound to the cell membrane and with the DNA being confined inside the nucleus in eukaryotes.
Moreover, the cytoplasm of cells is highly crowded, meaning that up to $40\%$ of the volume is occupied by macromolecules \cite{Luby-Phelps2000,Schnell2004}, while individual species are present only at very low concentrations.
On membranes crowding is even more severe \cite{Krapf2015}, due to attaching actin filaments creating static barriers \cite{Jin2007,Medalia2002}.
This macromolecular crowding leads to three excluded volume effects: (i) it forces moving molecules to diffuse around obstacles, or "crowders", slowing down the diffusion; (ii) it either decreases (diffusion limited case) or increases (reaction rate limited case or dimerizations) the reaction rates; and (iii) due to the impeded diffusion, it increases the spatial heterogeneity inside the cells, leading to spatial self organization \cite{Berry2002, Hansen2015}.
However, the effect of macromolecular crowding has not yet been fully understood, since single molecule tracking inside cells is still difficult.
Consequently, mathematical models are a crucial tool to understand reaction-diffusion processes in the intracellular environment.

One class of mathematical models are so called particle based reaction-diffusion models (PBRD), where we resolve individual molecules by following their diffusive paths and by modeling reactions as random events.
These are considered as the microscopic modeling level in reaction-diffusion simulations.
One group of PBRD are Brownian dynamics (BD) simulations, where the diffusive motion is modeled as a continuous time continuous space stochastic process.
When the particles are modeled as hard spheres these simulations inherently account for the excluded volume effects.
But the catch with these methods is, that they become computationally very expensive due to the high number of collision events when applied to dense environments.
An approximation to BD simulations are lattice based approaches, such as cellular automata (CA), where we still follow individual trajectories, but the possible particle positions are restricted to an artificial grid.
These methods are cheaper to simulate and in this paper we will investigate what influence the artificial lattice has on the excluded volume effects for the reaction and diffusion rates and the stochastic noise.
We therefore simulate both models in a two-dimensional plane representing cellular membranes.

This microscopic modeling framework captures the internal end external noise, that often play an important role in computational systems biology.
Reaction-rate equations (RREs) are a macroscopic approximation to this microscopic description.
These are deterministic ordinary differential equations describing the concentrations of the reacting molecules, and they are applicable when the law of large numbers holds, meaning that the molecules are present at large copy numbers, and when they are well-mixed in a dilute system.
They consequently do not capture the stochastic or space dependent effects inside cells, but provide suitable reference solutions in dilute media to compare the effect of macromolecular crowding to.

In the next section we will present the on- and off-lattice microscopic tools in more detail.
We will then perform simulations in environments with various crowder densities for static crowders and show that the diffusion is more severely slowed down when simulated in continuous than in discrete space.
In Section~\ref{sec:Reactions} we will extend the simulations to reactions in a crowded environment, and find that they are non-linearly affected by the artificial lattice.
We finally draw conclusions on the agreement between the BD and CA models when excluded volume effects are important in Section~\ref{sec:Conclusions}.

\section{Particle based reaction-diffusion models}
Particles in solution move due to their thermal energy, and the collisions with the smaller solvent molecules result in a random walk often modeled by Brownian motion.
This causes them to diffuse from high to low concentrations.
We will now present how to create sample trajectories of a reaction-diffusion system with on- and off- lattice simulations, where the particles diffuse with diffusion coefficient $\gamma_0$.
We demonstrate how to treat reaction events using the example of the association event
\begin{equation}
A+B\xrightarrow{k_A} C,
\label{eq:ForwardReaction}
\end{equation}
where $k_A$ is the intrinsic reaction rate.

\subsection{Continuous space}
Brownian dynamics (BD) simulations sample trajectories of Brownian motion in continuous space, or off-lattice.
To simulate a trajectory one can discretize time with a small enough fixed time step $\Delta t$.
We then draw a normally distributed random number $\xi\sim\calN(0,1)$ to compute the new particle position as
\begin{equation}
x(t+\Delta t) = x(t)+\sqrt{2\gamma_0\Delta t}\xi,
\end{equation}
and equivalently for the other space coordinates, since Brownian motion is independent along the Cartesian axes.
This algorithm is implemented among others in the freely available software package Smoldyn \cite{Andrews2010, Andrews2004}.
Here, a time step dependent binding radius is chosen such that the simulated reaction rate equals $k_A$.
Within this binding radius two reaction partners react with probability one and the method has been used for simulations in a crowded environment \cite[Ch.~4]{Gilbert2015}.
Similarly, in MCell \cite{Stiles1996} the position of a diffusing molecule after the discrete time step $\Delta t$ is sampled from the probability distribution of its position. 

Another algorithm for simulations in dilute media is Green's functions reaction dynamics (GFRD) \cite{VanZon2005}, where single particles or particle pairs are surrounded by protective domains.
An asynchronous time step is chosen in an event driven algorithm to be the time when the first particle leaves its protective domain and in this way one avoids simulating all the relatively uninteresting jumps between collision events.
%Reactions are here modeled by a reactive boundary condition for the pair propagator with parameter $k_A$.
An exact implementation is the so called first-passage kinetic Monte Carlo (FPKMC) algorithm \cite{Donev2010a,Oppelstrup2009,Takahashi2010}, used in the software ECell \cite{Tomita1999}.
The GFRD algorithms, however, become computationally very expensive \cite{Lee2008}, when applied to non-dilute systems and we will perform BD simulations with the software Smoldyn to investigate the excluded volume effects in continuous space.

Volume exclusion effects are inherently simulated with these algorithms \cite{Takahashi2005}, when the particles are represented as hard spheres rather than point particles. The review article \cite{Schoneberg2014}, summarizes which of the available software packages allow for this feature. 

\subsection{Discrete space}
To gain further efficiency when simulating PBRD we will now restrict the particles' trajectories to an artificial lattice.
These so called cellular automata (CA) or lattice gas automata models \cite{Boon1996} are widely used to investigate excluded volume effects on both the diffusion rates \cite{Ellery2015} and the reaction rates \cite{Berry2002,Grima2006,Schnell2004}.
First, the domain is discretized into a grid or lattice.
Each lattice site can then contain at most one molecule and at each time step the molecules are picked in random order to jump to a neighboring site.
If the sampled target site is occupied, either a reaction happens with probability $p_A$ or the move is rejected, which models the excluded volume effects.
Since the mean square displacement (MSD) of a diffusing molecule in $d$ dimension is 
\begin{equation}
\langle\fatx^2(t)\rangle=2d\gamma_0t
\label{eq:MSD}
\end{equation}
we choose the time step
\begin{equation}
\Delta t = \frac{h^2}{2d\gamma_0}
\label{eq:TimeStep}
\end{equation}
for a lattice spacing $h$.
The algorithm for the simple reaction \eqref{eq:ForwardReaction} until final time $T$ then reads as follows.
\begin{algorithm}[h!]
				\caption{Cellular Automata}
				\label{alg:CA}
				\begin{algorithmic}[1]
				\State Initialize system by placing initial numbers of $A$, $B$ and $C$ molecules randomly on the grid.
				\While {$t<T$}
				\State Choose molecules in random order.
				\For {each molecule} 
				\State Randomly choose a nearest neighbor site as target.
				\If{target site is empty} 
					\State Move molecule.
				\Else
					\If{ molecule is $A$($B$) and target is occupied by $B$($A$)}
						\State Generate a random number $\xi$.
						\If{$\xi<p_A$}
							\State Replace $A$ and $B$ with a $C$ molecule at target site.
						\Else \State Reject the jump.
						\EndIf
					\Else \State Reject the jump.
					\EndIf
				\EndIf
				\EndFor
				\State Update $t:=t+\Delta t$.
				\EndWhile
				\end{algorithmic}
				\end{algorithm}

This algorithm has been used for Cartesian and  hexagonal lattices and in \cite{Grima2006} it is shown that the algorithm is sensitive to the choice of grid in a crowded environment and in general overestimates the excluded volume effects on the reaction rates as compared to BD simulations.
In \cite{Ellery2015} the algorithm has been extended to model different molecule shapes, which consist of combinations of Cartesian lattices, and the diffusivity of moving molecules among static crowders is computed for a mix of shapes. 

In the next sections we will perform both, BD and CA simulations in crowded environments to examine if CA are a good approximation of BD in this setting.
The crowders are here represented as static and inert particles, that do not actively participate in the reaction processes.

\section{Diffusion simulations}
\begin{figure}[h!]
\centering
\subfigure[Cartesian]{\includegraphics[width=.19\textwidth]{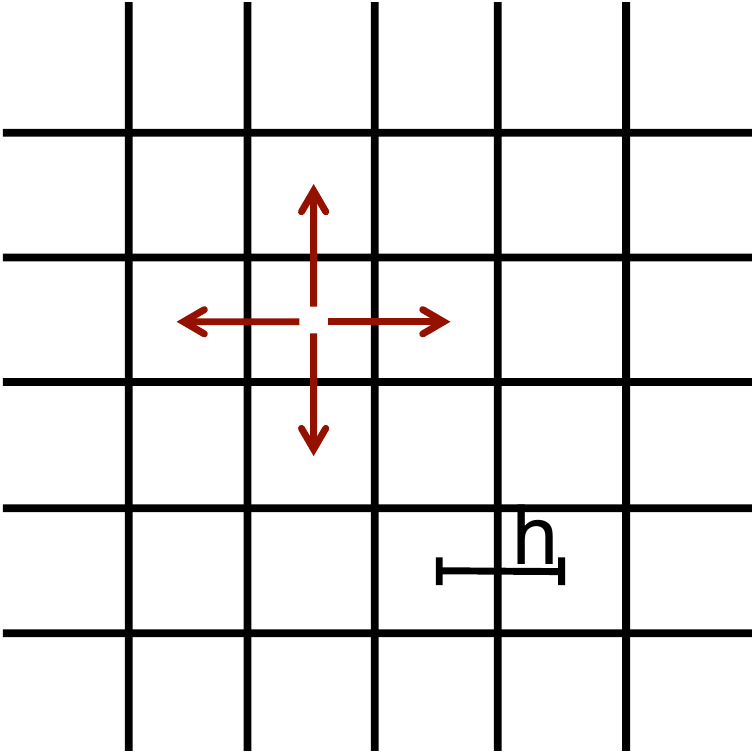} }\hspace{1.3cm}
\subfigure[Diagonal]{\includegraphics[width=.19\textwidth]{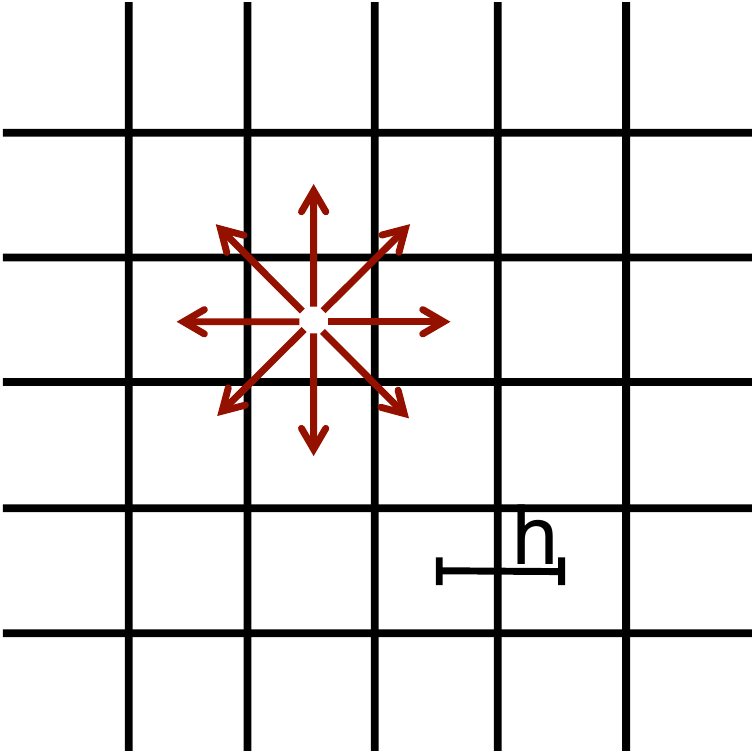} }\hspace{1.3cm}
\subfigure[Hexagonal]{\includegraphics[width=.19\textwidth]{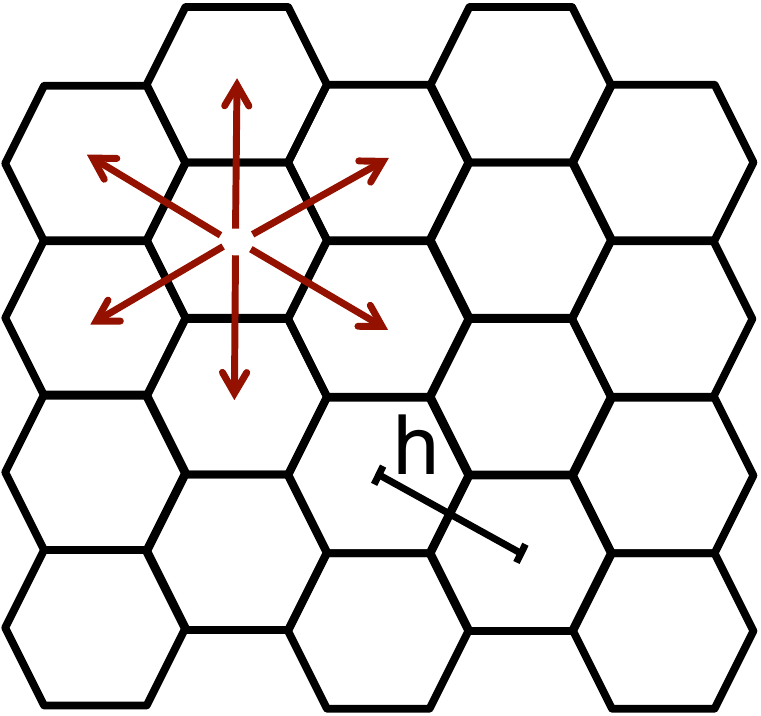} }
\caption{Lattices for CA simulations with different degrees of freedom (dof). (a) Cartesian mesh with 4 dof. (b) Cartesian mesh with 8 dof. (c) Hexagonal mesh with 6 dof.}
\label{fig:Lattices}
\end{figure}

In all the experiments we choose $\gamma_0=1$ and perform the on-lattice simulations according to Algorithm~\ref{alg:CA}, referred to as CA, and use the open source software Smoldyn for the off-lattice simulations, referred to as BD.
We investigate three different lattices (Cartesian, Cartesian with diagonal jumps and hexagonal) in the CA simulations, see Fig.~\ref{fig:Lattices}, where the coefficients for the diagonal jumps are chosen as in \cite{Meinecke2016}.
We discretize the two-dimensional domain $50\times 50$ with $h=1$, so that all molecules (crowders and diffusing) have size $1$ in the CA simulations and are spheres with $r=0.5$ in the BD simulations.
The resolution of the on-lattice simulations is defined by $h$ and according to \eqref{eq:TimeStep} this leads to the CA time step $\Delta t^{CA}=0.25$.
For the BD simulations we choose a finer resolution with $\Delta t^{BD}=0.001$, to guarantee that the off-lattice simulations are time step independent.
In order to investigate the diffusive motion over long times without encountering boundary effects we implement periodic boundary conditions on all boundaries.

\begin{figure}[t!]
\centering
\subfigure[$\phi=0.2$]{\includegraphics[width=.31\textwidth]{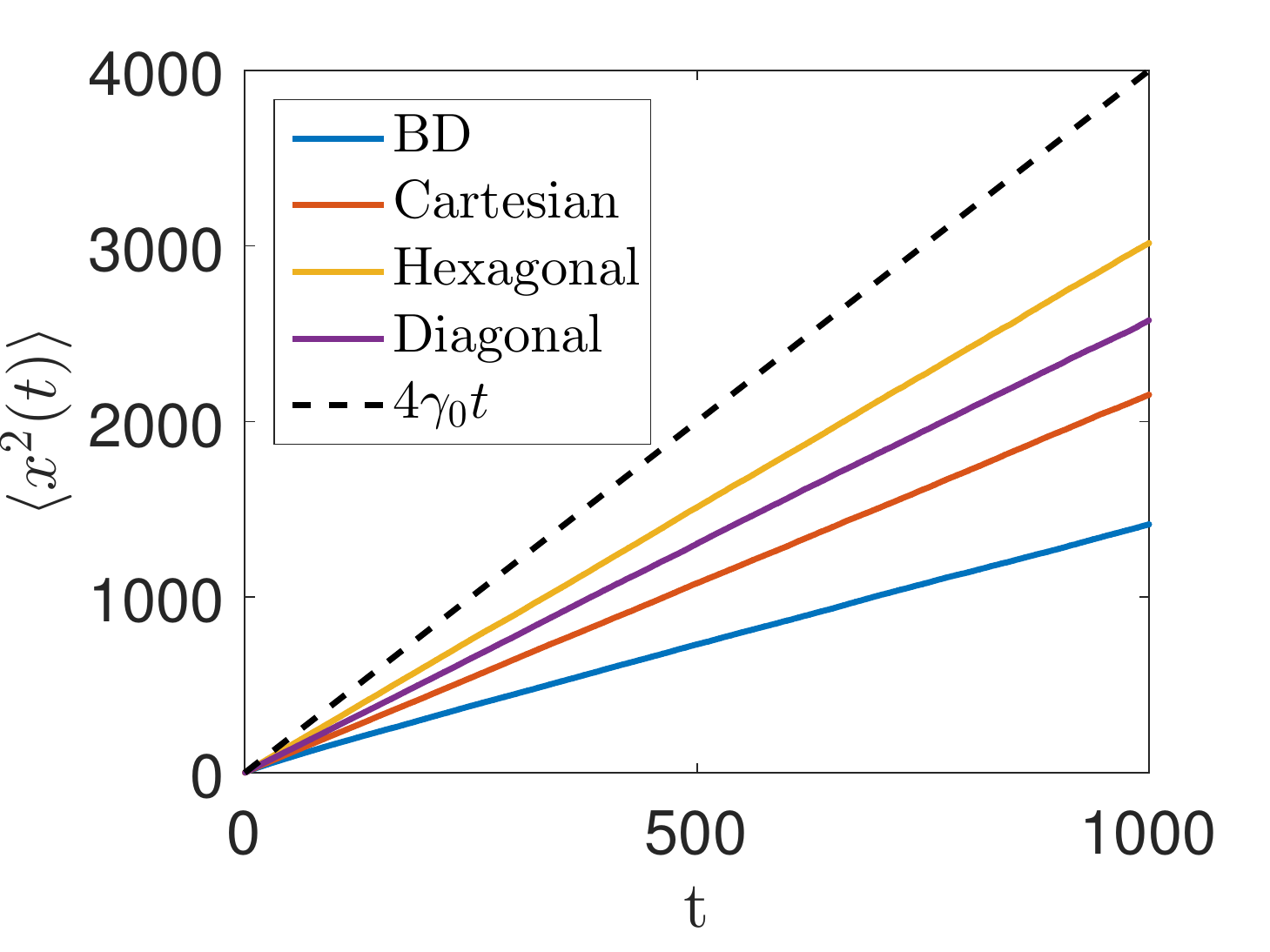} }
\subfigure[$\phi=0.4$]{\includegraphics[width=.31\textwidth]{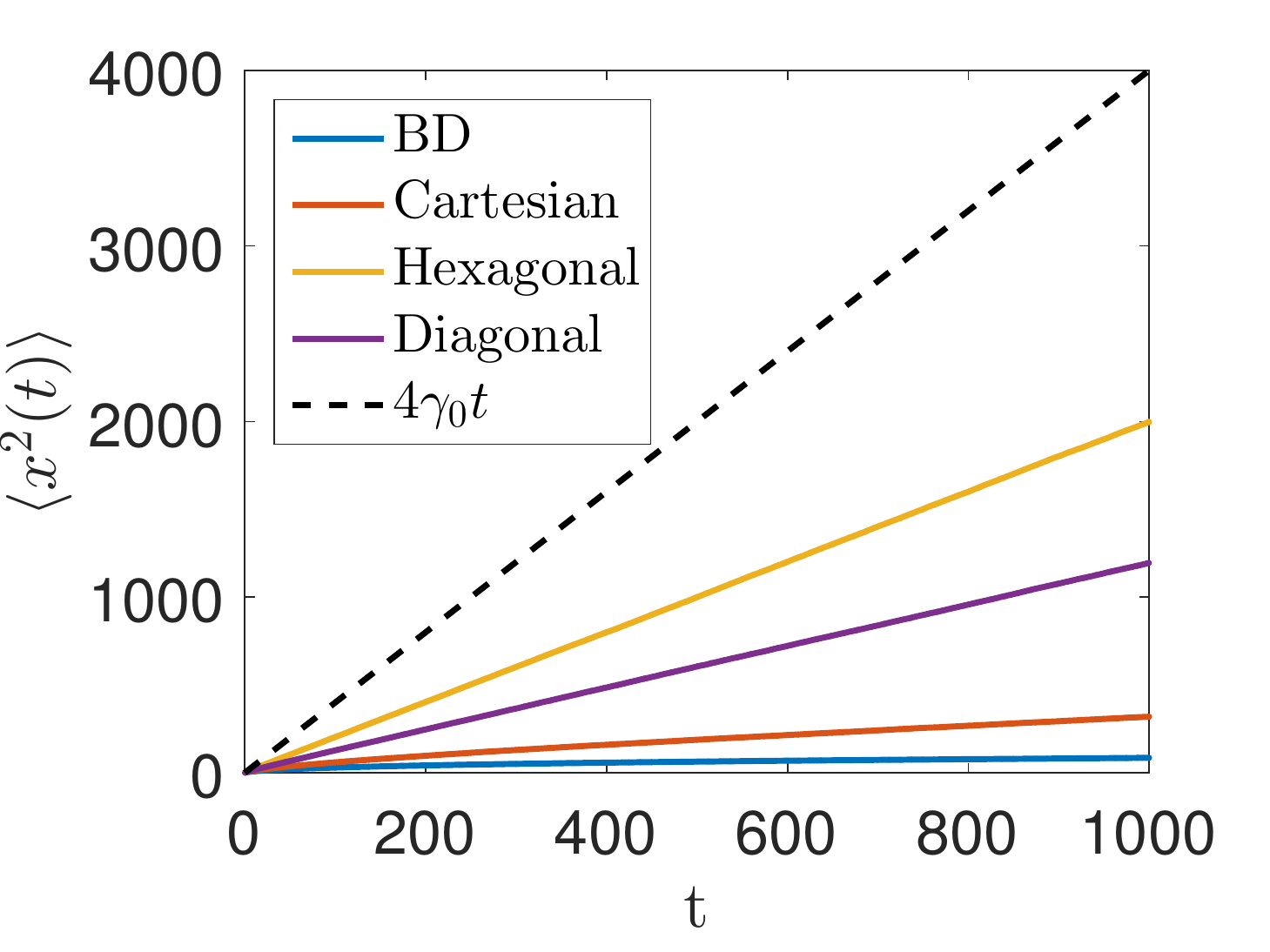} }
\subfigure[$\phi=0.6$]{\includegraphics[width=.31\textwidth]{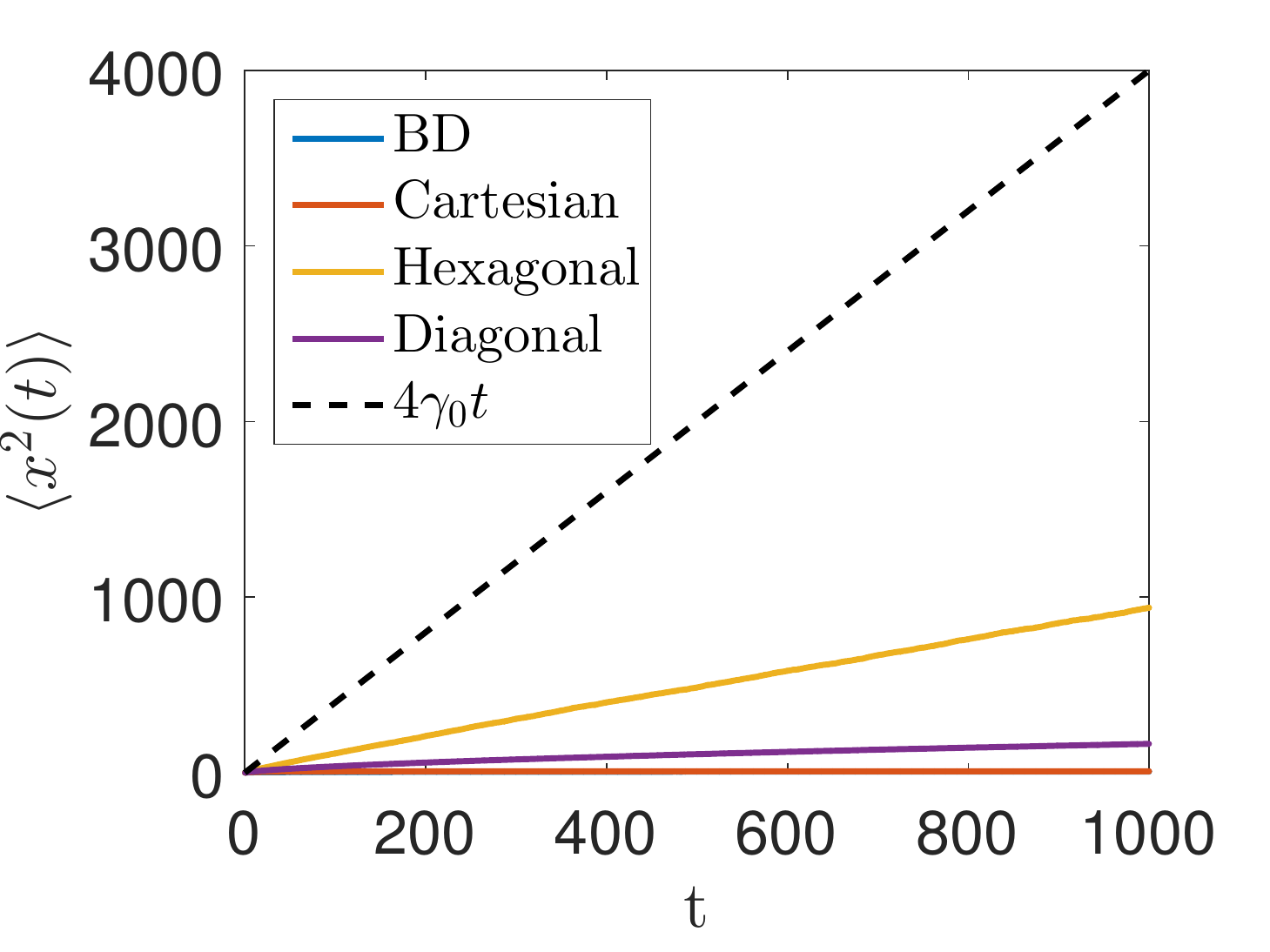} }\\
\subfigure[$\phi=0.2$]{\includegraphics[width=.31\textwidth]{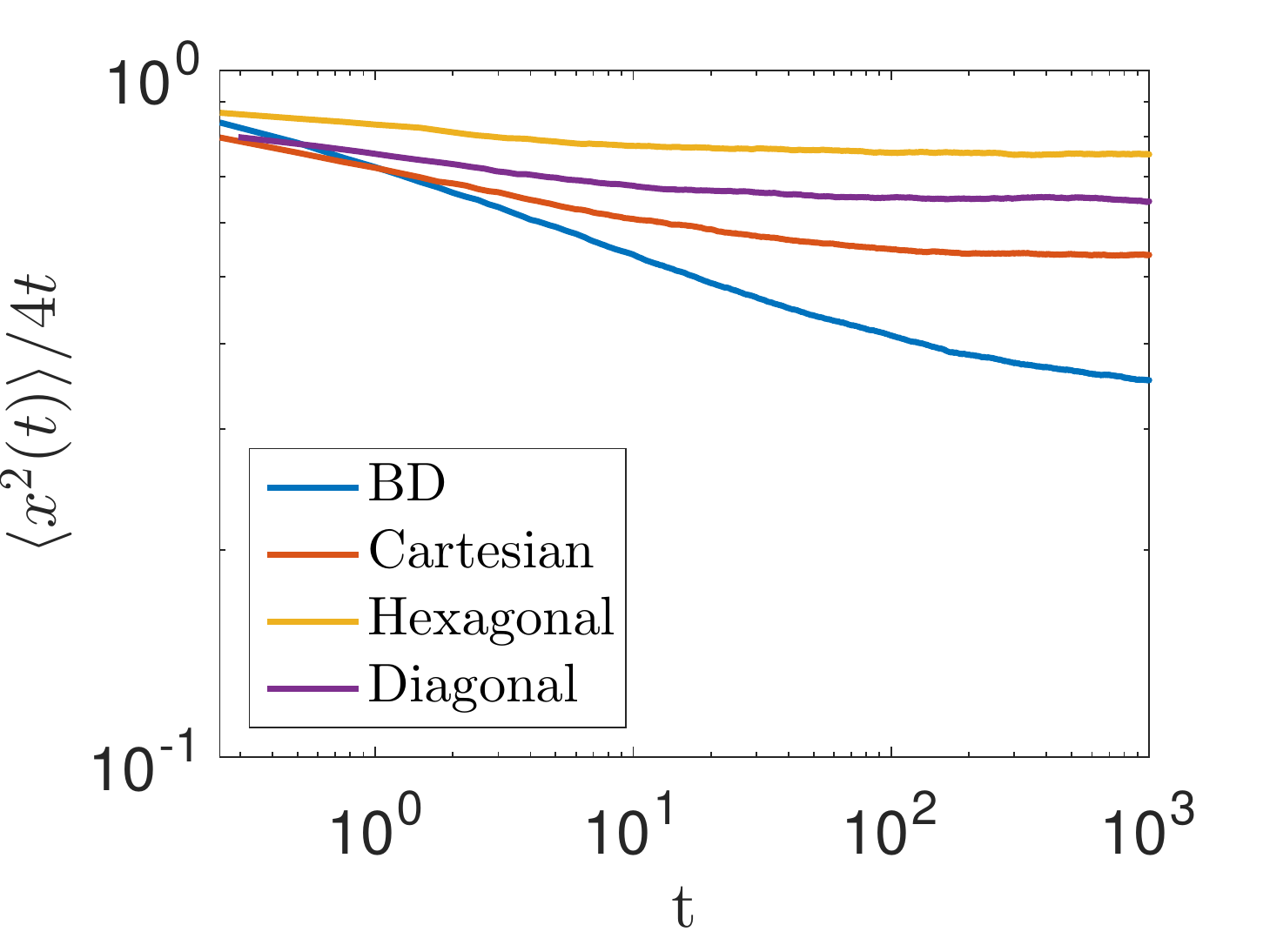} }
\subfigure[$\phi=0.4$]{\includegraphics[width=.31\textwidth]{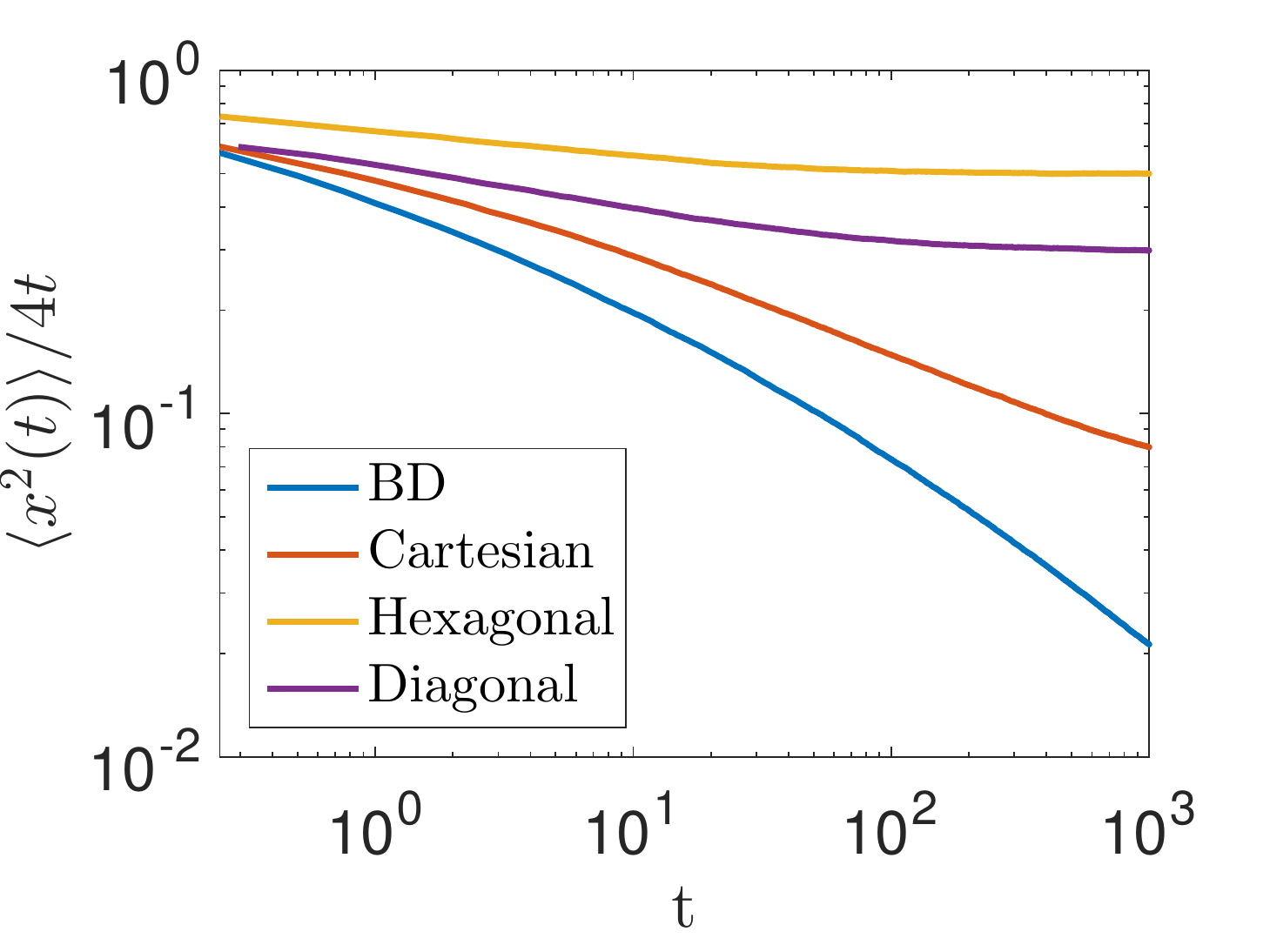} }
\subfigure[$\phi=0.6$]{\includegraphics[width=.31\textwidth]{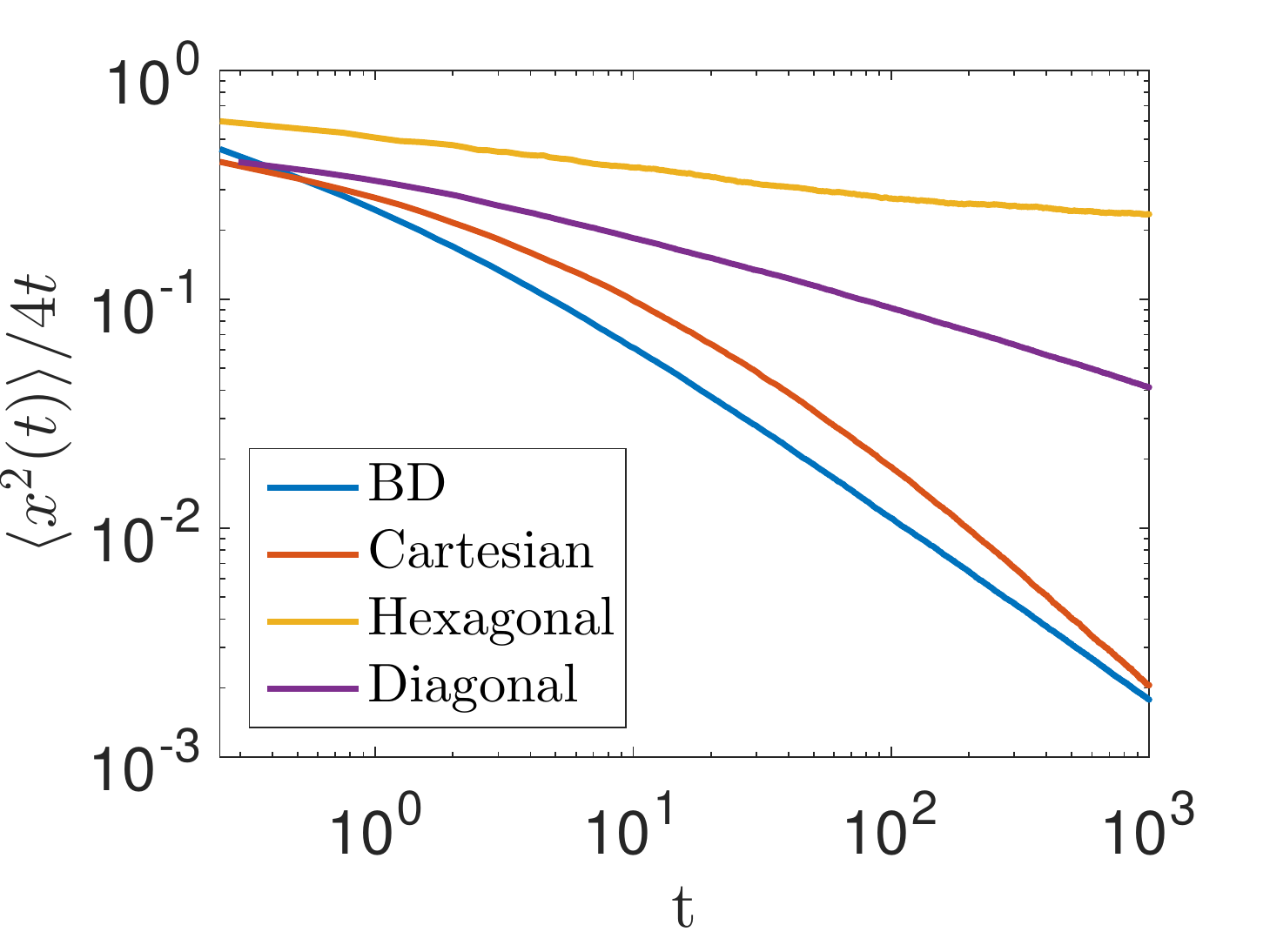} }
\caption{The MSD for a diffusing molecule, where $\phi$ is the fraction of occupied volume of static crowders. First row: $\langle\fatx^2(t)\rangle$. The reference line is the MSD in dilute medium \eqref{eq:MSD}. Second row: $\langle\fatx^2(t)\rangle/4t$. Increasing $\phi$ here leads to increasingly slower diffusion and increases the difference between the models, where excluded volume effects are strongest in the off-lattice simulations.}
\label{fig:MSDstatic}
\end{figure}

The MSD of a diffusing molecule surrounded by static crowders is plotted in Fig.~\ref{fig:MSDstatic} for an increasing fraction of occupied volume $\phi$.
In CA we simulate $10^5$ trajectories each in a different crowder distribution and in BD we simulate $100$ different crowder distributions with each $1000$ trajectories.
Note that the more accurate BD simulations with the small time step $\Delta t^{BD}$ run ca.\ 58 times slower than the CA simulations for $\phi=0.4$.

\begin{figure}[t!]
\centering
\subfigure[]{\includegraphics[width=.2\textwidth]{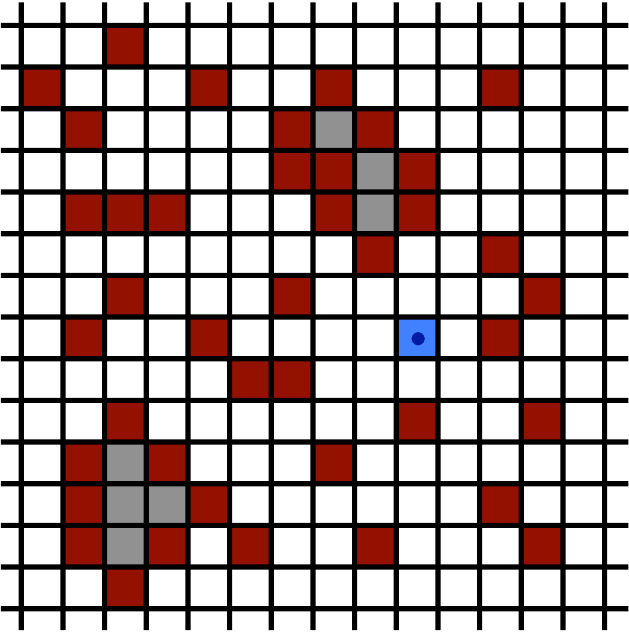} }\hspace{1cm}
\subfigure[]{\includegraphics[width=.2\textwidth]{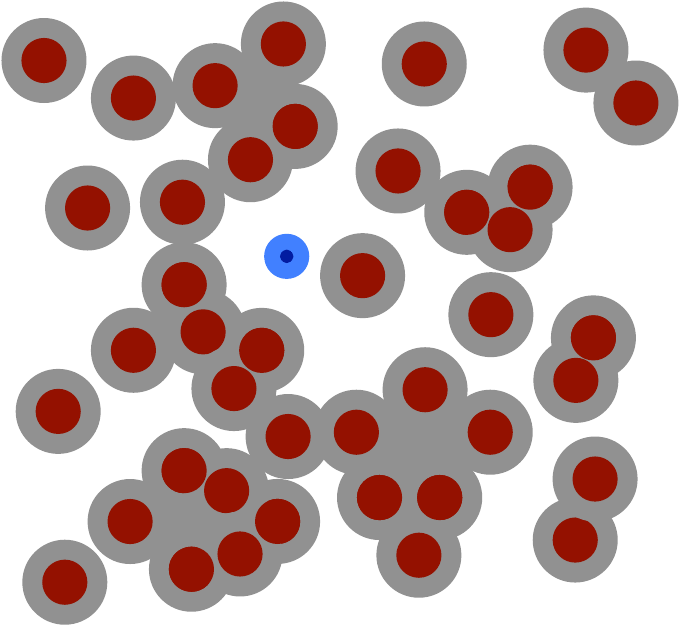} }\hspace{1cm}
\subfigure[]{\includegraphics[width=.2\textwidth]{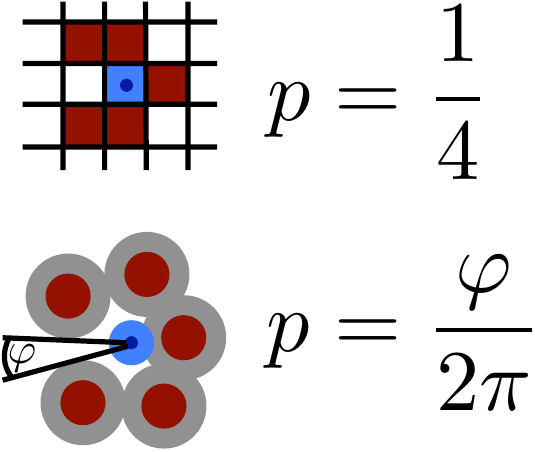} }
\caption{The diffusive motion is more obstructed by crowding molecules when simulated off-lattice, than on-lattice. This can be understood when comparing the effectively excluded volume (grey) caused by the crowding molecules (red) for the center of mass (dark blue spot) of the moving molecule (blue). (a) CA on a Cartesian lattice, where the blue particle is rather mobile. (b) Off-lattice distribution of crowders as in BD resulting in the blue particle being almost confined to the small center area. (c) The probability $p$ that the possible direction of movement is chosen.}
\label{fig:Block}
\end{figure}

If no crowders are placed in the system ($\phi=0$) all models agree with the theoretical MSD in dilute medium \eqref{eq:MSD} (plot not shown).
But, for an increasing $\phi$ the on- and off-lattice models increasingly differ and the speed of diffusion decreases, which is expected.
The unexpected finding is that the diffusion is more heavily slowed down with BD than with CA simulations.
This appears counter-intuitive at first, since the artificial grid decreases the available directions of movement and the off-lattice model with infinitely many degrees of freedom (dof) is expected to simulate more mobile particles.
But, the lattice also orders the particles, so that they effectively exclude less space, see Fig~\ref{fig:Block}(a)$\&$(b).
This principle can be understood intuitively when considering a parking lot with predefined parking spots, imagine finding a spot or leaving the lot if all cars were parked randomly.
The restriction of the degrees of freedom also makes it more probable to choose the possible direction out of a finite number of lattice directions as compared to the probability to sample a jump in the small angular direction $\varphi$, see Fig.~\ref{fig:Block}(c).
The increased flexibility on the hexagonal lattice leads to even faster diffusion than on a Cartesian, but allowing for diagonal jumps does not increase diffusion any further, since most of the jumps ($82\%$) are sampled along the Cartesian axes. 
Hence, the Cartesian grid with only 4 dof has the closest agreement with the off-lattice simulations with infinitely many dof, contrary to the findings in \cite{Grima2006} for fractal-like reaction kinetics.
For very high and unbiological crowder densities ($\phi=0.6$), particles in BD and CA simulations on a Cartesian mesh effectively no longer move and only the on-lattice models with more dof allow the particles to find a free passageway.

\begin{figure}[h!]
\centering
\includegraphics[height=.16\textheight]{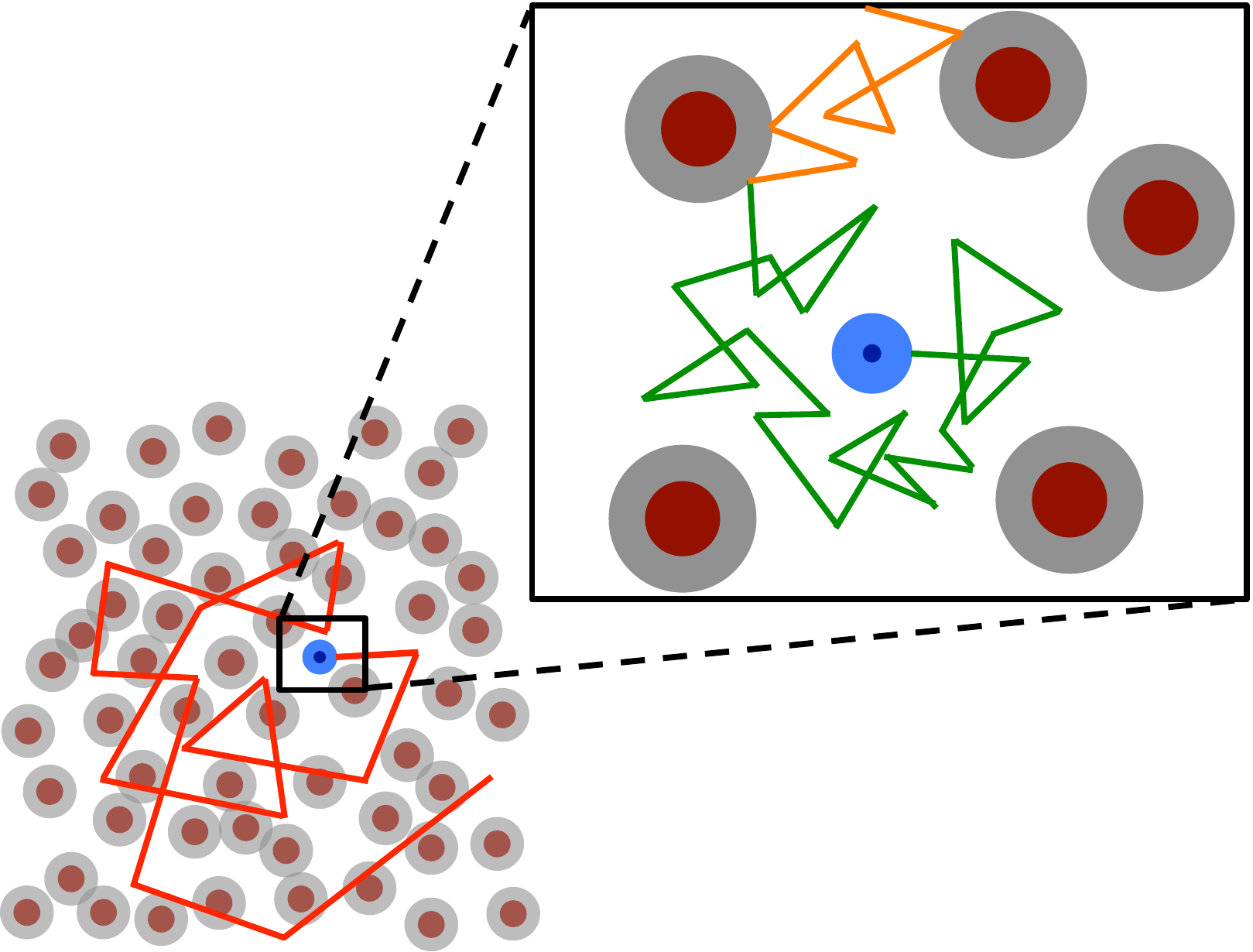}\hspace{.8cm}
\includegraphics[height=.18\textheight]{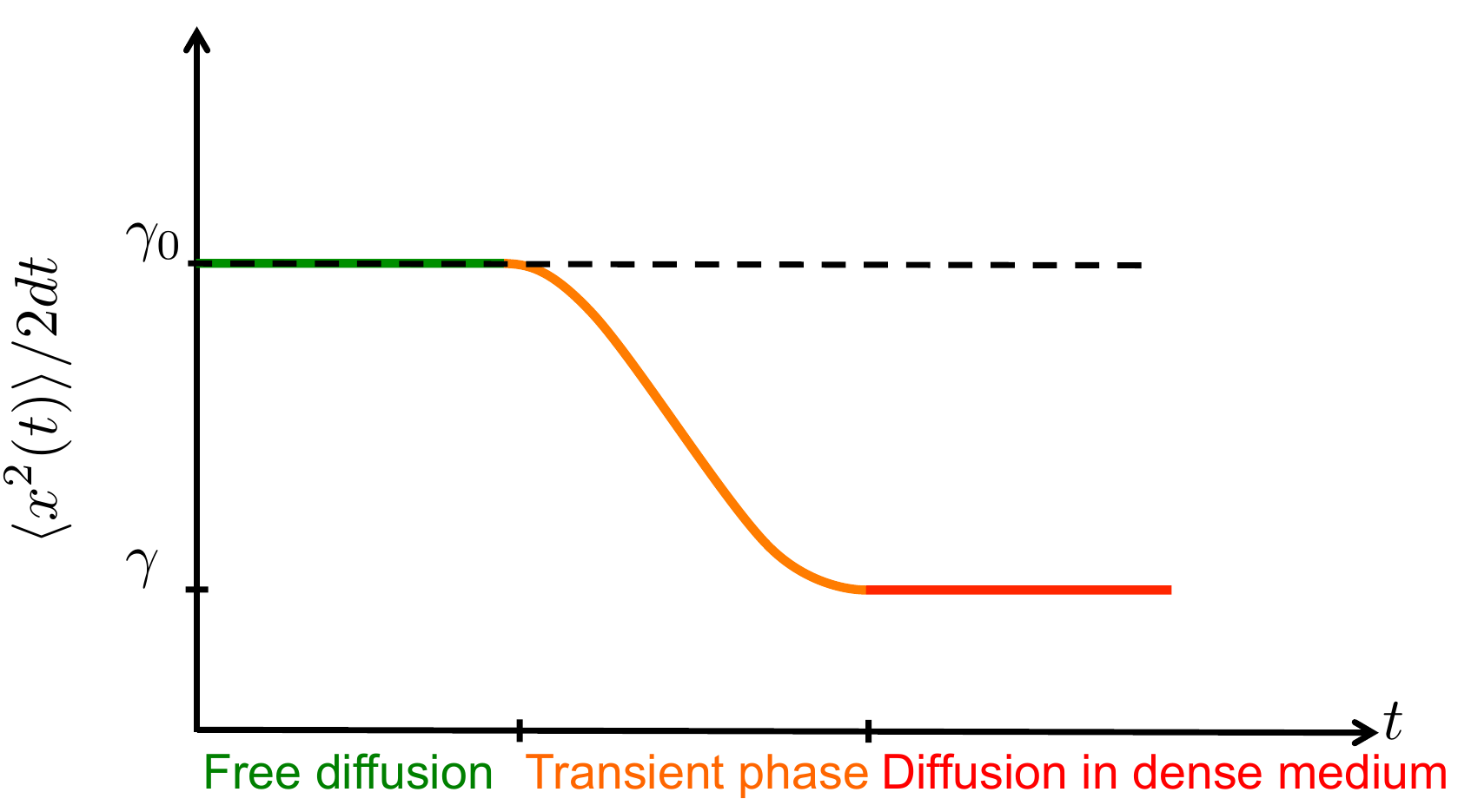}
\caption{(a) Diffusion in an environment with static crowders. Before the diffusing molecule encounters the first obstacles it moves with the dilute diffusion speed $\gamma_0$ (green). When it starts colliding the diffusion it slowed down (orange), until an average slower diffusion is observed on long time scales (red). (b) The anomalous behavior of the MSD (solid line) and the constant behavior of diffusion in dilute medium (dashed line).}
\label{fig:Anomalous}
\end{figure}

For normal diffusion in a dilute medium in integer dimension $d$ the MSD grows linearly with time \eqref{eq:MSD}.
In the second row in Fig.~\ref{fig:MSDstatic} we further observe that the MSD is no longer linear, but that diffusion becomes anomalous with a time varying diffusion constant.
This behavior is explained in \cite{Krapf2015} and illustrated in Fig.~\ref{fig:Anomalous}, where the change in the diffusion constant is due to encountering the surrounding obstacles, before converging to an average long time behavior.
Note, that we do not depict the small solvent molecules causing the random walk.
However, when the crowder density is higher than the percolation threshold, the obstacles form a cluster traversing the whole domain, so that disconnected subdomains form and space has a non-integer dimension smaller than two, consequently the MSD is non-linear for all times \cite{benAvraham2000,Weigel2012}.
The percolation thresholds for lattices are $\phi=40.73\%$ (Cartesian) and $\phi=50.30\%$ (Hexagonal) \cite{Newman2000}. 
For the off-lattice case with partially overlapping disks for the excluded volume as illustrated in Fig.~\ref{fig:Block}(b) it is more difficult to find exact values, but if the free space would consist of fully overlapping spheres $67.63\%$ \cite{Quintanilla2000} would have to be freely available which would leave us with an occupancy fraction of $32.37\%$ for the effective excluded volume.
In the second row in Fig.~\ref{fig:MSDstatic} we see that the simulations reveal sustained anomalous diffusion for the cases when the percolation threshold has been exceeded.

\begin{figure}[h!]
\centering
\includegraphics[width=.45\textwidth]{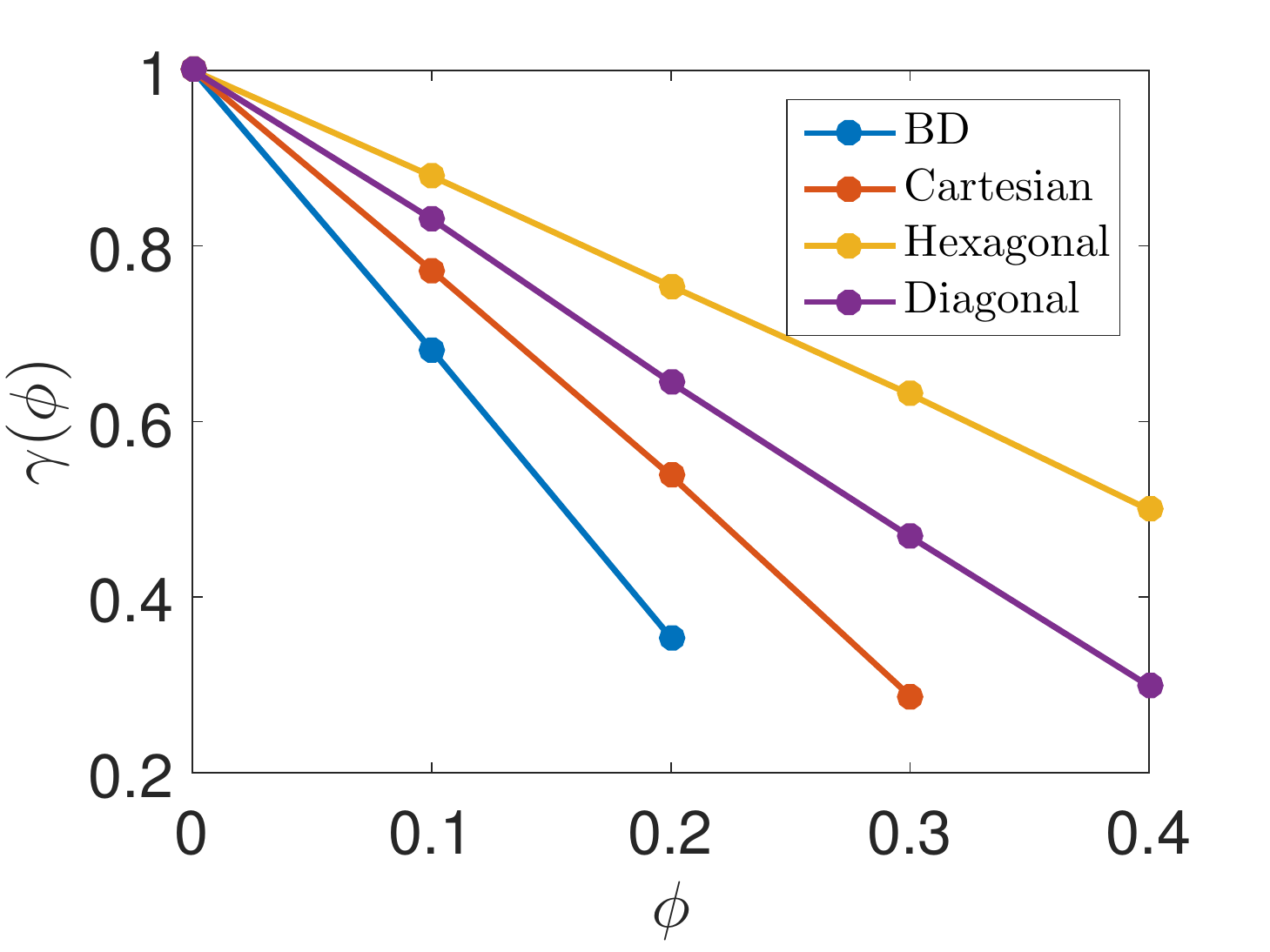} 
\caption{The effective long time diffusivity $\gamma(\phi)$ obtained from particle based on and off-lattice simulations. Due to the random distribution of crowders in BD, the diffusivity is slowest in the off-lattice simulations and increases with an ordering grid and more possible jump directions.}
\label{fig:Diffusivity}
\end{figure}

To quantify the excluded volume effects on the effective diffusivity, we evaluate the diffusion constant $\gamma(\phi)$ after the transient phase by taking the last value of the plot $\langle x^2(t)\rangle/(4t)$, for the cases converging to normal slower diffusion, see Fig.~\ref{fig:Diffusivity}.
As expected, the effective diffusion constant $\gamma(\phi)$ decreases with increasing crowder density and we can clearly observe the lattice artifact of underestimating the excluded volume effect on the diffusivity.
It also appears that $\gamma(\phi)$ depends linearly on $\phi$ for all models, and the decrease in diffusivity for BD agrees with the findings in \cite{Meinecke2016b} for equally sized spheres.

\begin{figure}[h!]
\centering
\subfigure[BD]{\includegraphics[width=.45\textwidth]{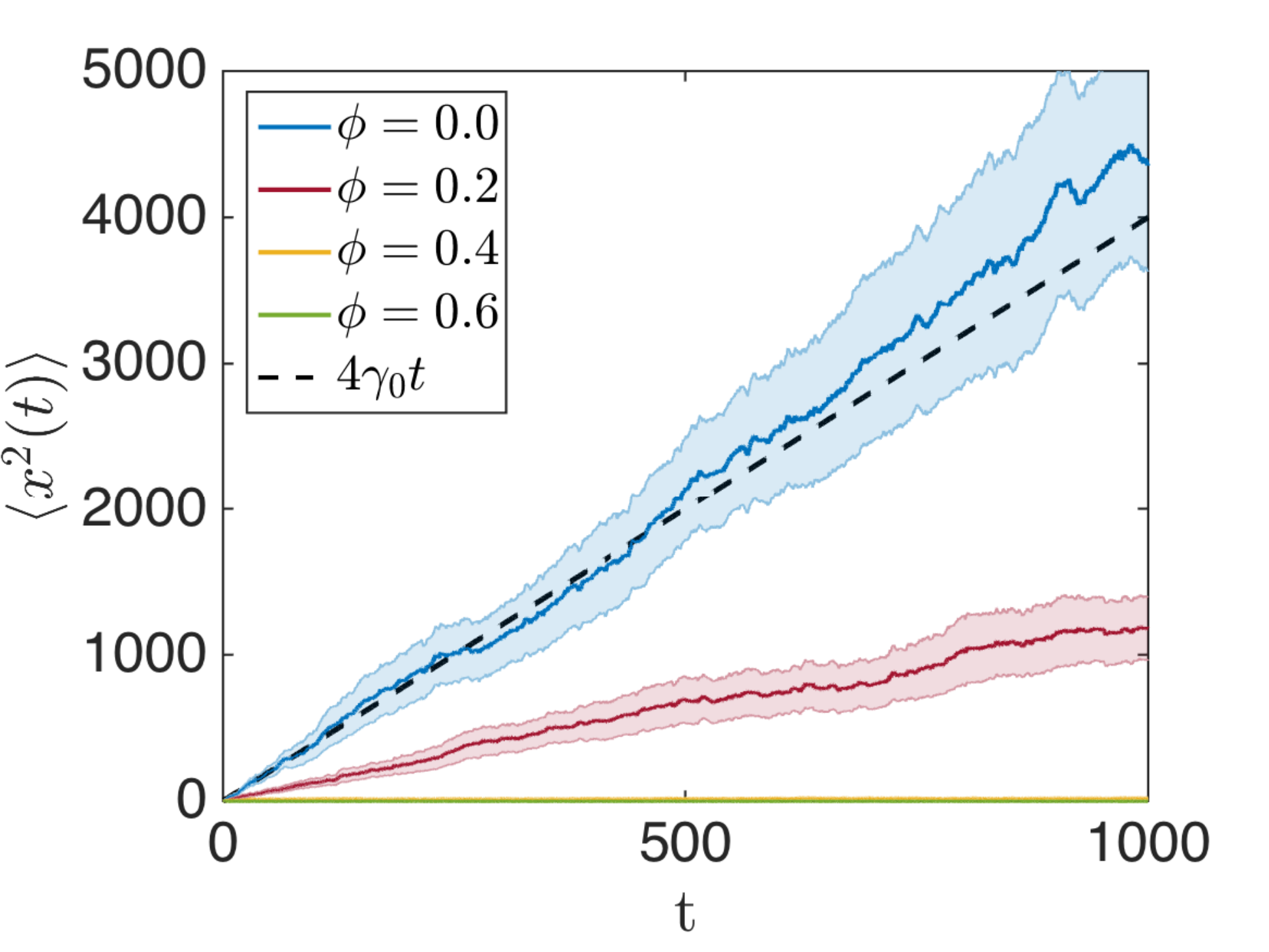} }
\subfigure[Cartesian]{\includegraphics[width=.45\textwidth]{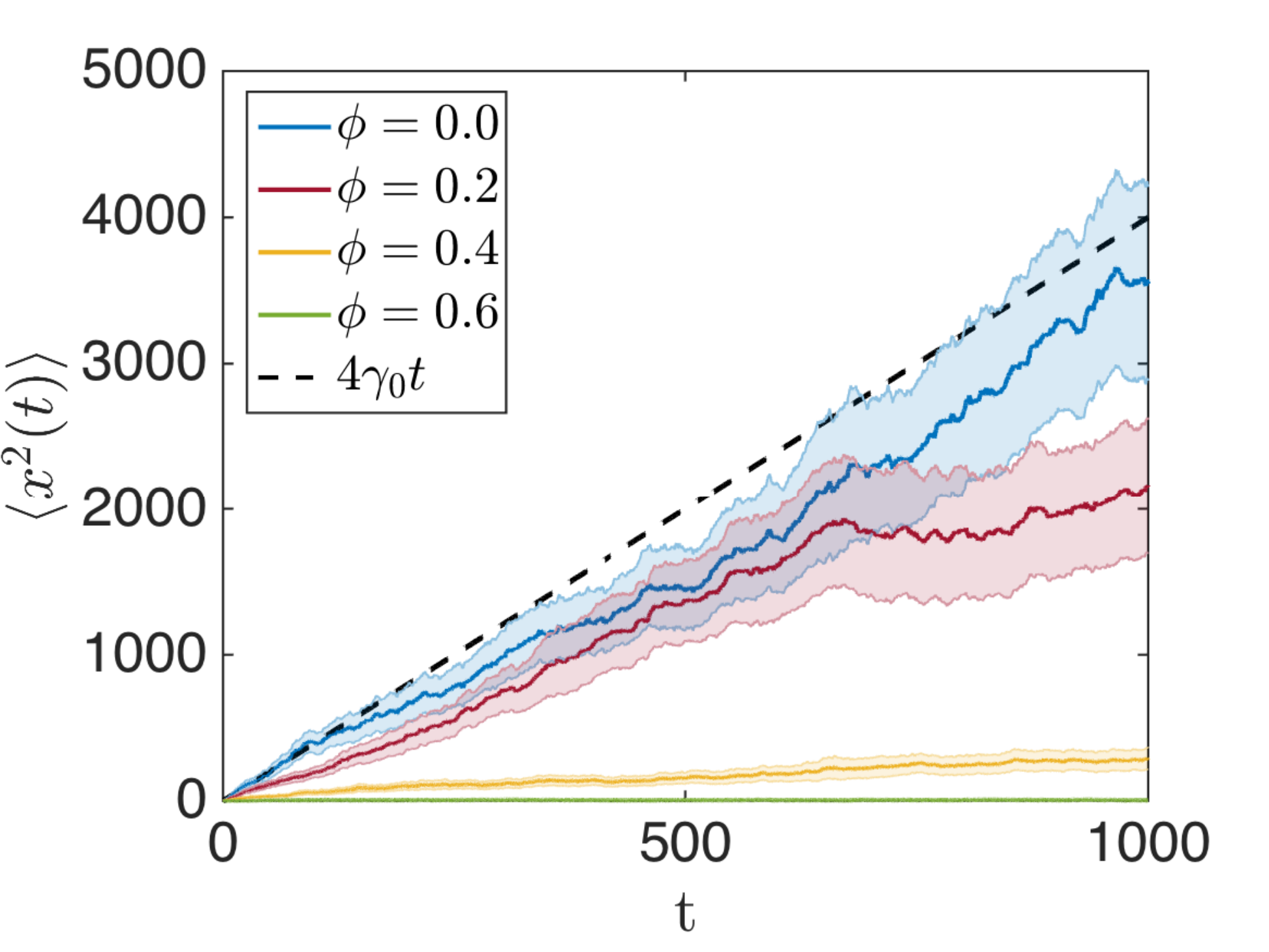} }\\
\subfigure[Hexagonal]{\includegraphics[width=.45\textwidth]{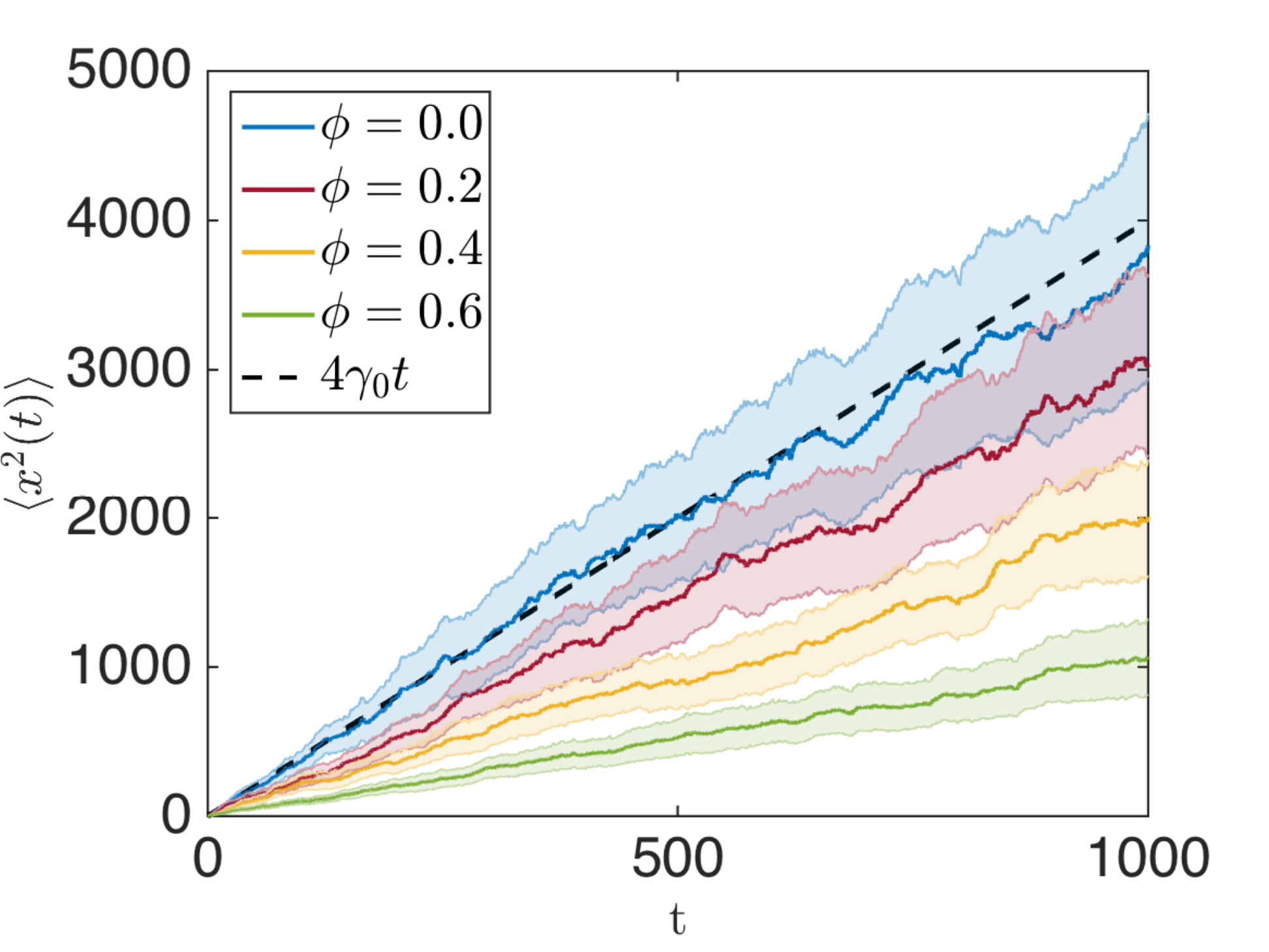} }
\subfigure[Diagonal]{\includegraphics[width=.45\textwidth]{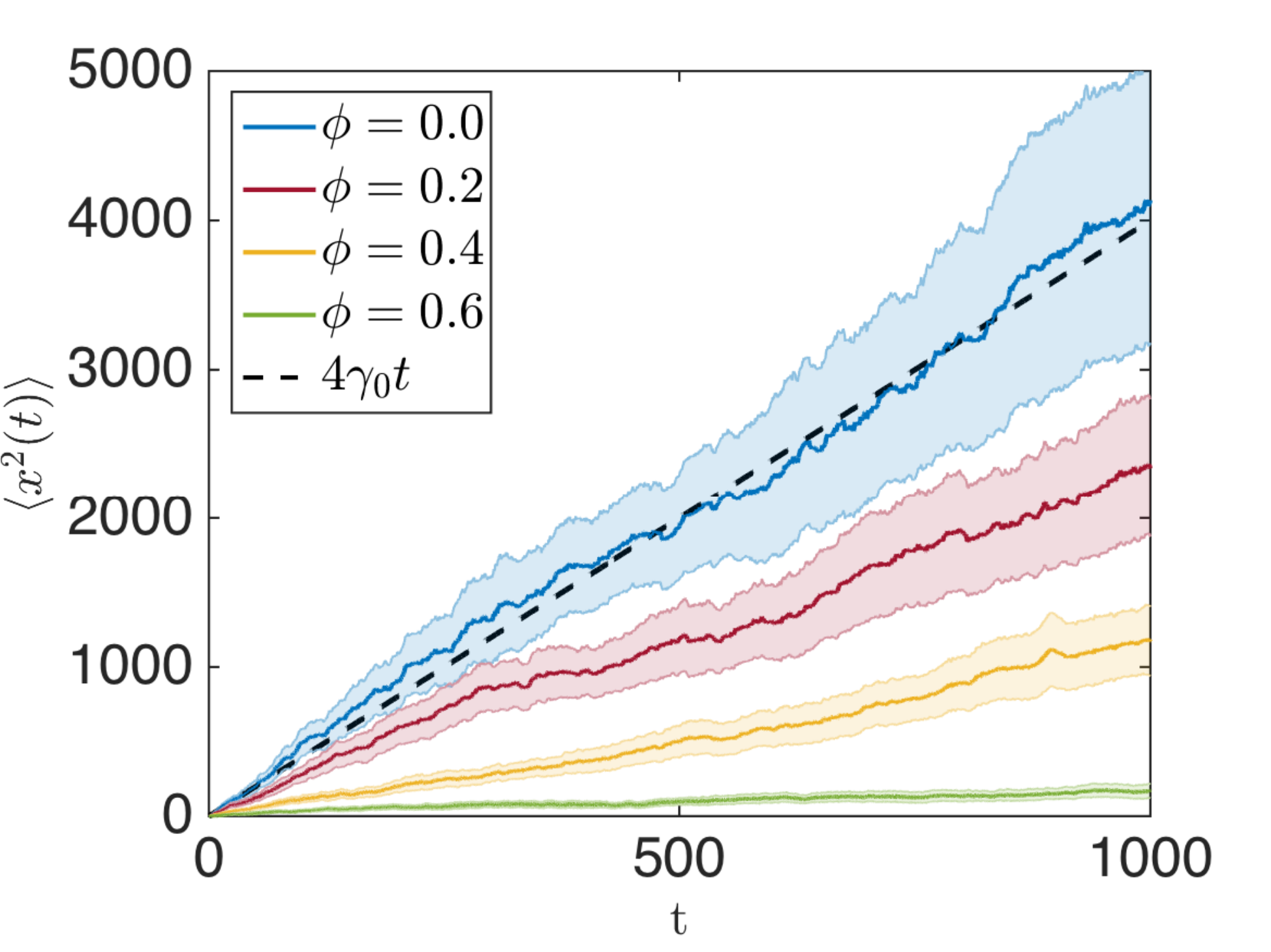} }
\caption{MSD and the $95\%$ confidence interval simulated for 100 trajectories in one crowder distribution. The variance decreases with increasing crowder density, as there is less space available for the molecules to diffuse in.}
\label{fig:MSD_STD}
\end{figure}

We now simulate 100 trajectories for one specific crowder distribution and plot the mean together with the $95\%$ confidence interval in Fig.~\ref{fig:MSD_STD}, in order to investigate the excluded volume effects on the variance of the diffusive motion.
As $\phi$ increases, less and less space becomes available for the particles' diffusion and hence the variance and diffusion speed decrease.
Another interpretation is that the higher number of particles in the system leads to a more deterministic behavior.
Similarly, the BD simulations resulting in slower diffusion have a smaller trajectory to trajectory variance, but we do not observe a grid effect for the variance otherwise.

\newpage
\section{Reaction-diffusion simulations}\label{sec:Reactions}
The excluded volume also has a thermodynamic effect on the reaction rates.
For a high diffusion constant the system can be considered well-mixed, and the rate of reaction rate limited reactions is increased since the obstacles decrease the reaction volume.
Moreover, dimers effectively exclude less space than two monomers, so that dimerization is also favored by crowding effects \cite{Hall2003}.
Diffusion-limited reactions on the other hand are impeded due to the slowed down diffusion and the increased time for the reaction partners to meet.
There exist many models for reactions in the crowded cell environment, such as fractional dynamics \cite{Grima2006,Schnell2004}, a power law approximation of the RREs \cite{Savageau1976}, and fractional \cite{Krapf2015} and multifractional Brownian motion \cite{Marquez-Lago2012}, or scaled particle theory (SPT) to compute the reaction rates using statistical physics \cite{Grasberger1986,Grima2010,Hall2003,Ridgway2008}.
To examine the excluded volume effects on the reaction rates and in particular if CA can capture the same effects as BD, we compare association, dissociation and reversible reactions
\begin{equation}
A+B\to C,\quad\quad C\to A+B,\quad\quad A+B\rightleftharpoons C,
\end{equation}
when simulated with BD and with CA.
All CA simulations are here performed on a Cartesian grid, since it shows the best agreement with off-lattice simulations in the pure diffusive case.
We model the complex $C$ in two ways: (i) $C$ has the same size as $A$ and $B$; (ii) $C$ has double the size of $A$ and $B$, meaning it is a $2\times 1$ molecule in CA and a sphere with radius $r_C=\sqrt{2}r_A$ in BD, see Table~\ref{tab:CModel}.
The latter model appears to be more realistic when we simulate a system where the molecules occupy volume, since no mass is lost in a binding reaction and the resulting complex occupies the same volume as the two reacting molecules together.
The restriction that molecules have to be composed of Cartesian voxels in CA, however, forces us to choose a rather unrealistic representation of C in that case.
To account for the larger size of $C$ we adjust its diffusion constant $\gamma_0^C=\gamma_0/\sqrt{2}$, and we extend the CA model to simulate particles of different sizes by adding an extra time step $\Delta t^{CA,C}=\sqrt{2}\Delta t^{CA}$ for the jumps of $C$.

In this section $a(t)$ denotes the number concentration of $A$ molecules
\begin{equation}
a(t)=\frac{A(t)}{V},
\end{equation} 
where $A(t)$ is the number of A molecules in the system with volume $V$ at time $t$.
We focus on diffusion-limited reactions by choosing the reaction probability to be one and compute $10^3$ sample trajectories in $10^3$ different crowder distributions and plot their mean values in all experiments.
The crowders are represented as an additional molecular species, which is static and inert and hence does not actively affect the reactions.

\begin{table}[h]
\centering
\begin{tabular}{c|ll}
& Model I & Model II \\
\hline\\
BD & \includegraphics[height=.06\textheight]{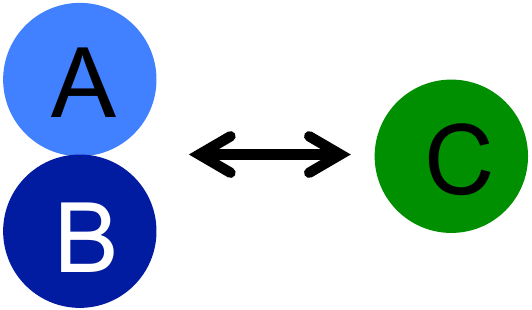} \hspace{.2cm} & \includegraphics[height=.06\textheight]{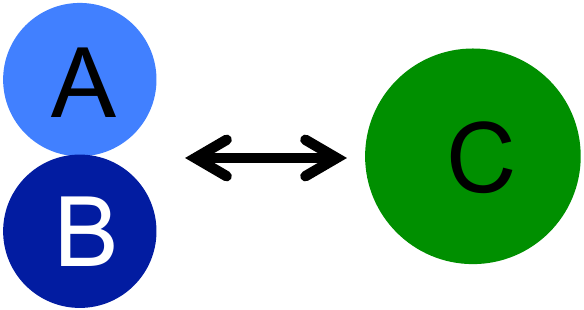} \\
\\
CA & \includegraphics[height=.06\textheight]{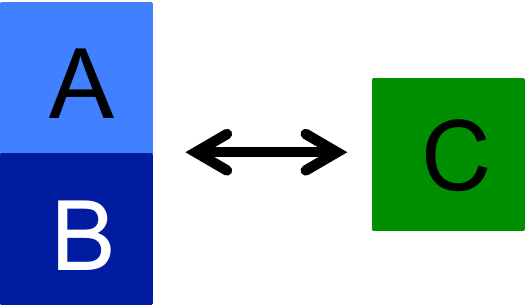} \hspace{.2cm} & \includegraphics[height=.06\textheight]{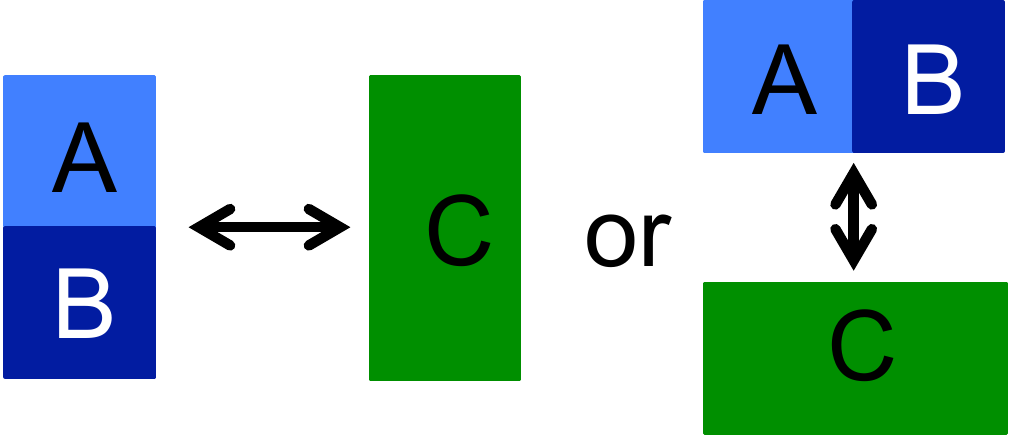}
\end{tabular}
\caption{We model the complex $C$ in two ways. Model I: C has the same size as $A$ and $B$. Model II: $C$ has double the size of $A$ and $B$.}
\label{tab:CModel}
\end{table}

\newpage
\subsection{Association events}
We first consider the bimolecular binding reaction
\begin{equation}
A+B\xrightarrow{k_A} C.
\end{equation}
When volume exclusion is taken into account the reaction radius in Smoldyn is not well calibrated \cite{Smoldyn}.
Hence, we first experimentally find the reaction rate $k_A^{BD}$ (Table~\ref{tab:BindingTime}) for Smoldyn, such that the mean binding time for two molecules $A$ and $B$ in the $[20\times 20]$ domain with periodic boundary conditions without any crowders agrees with that of a CA simulation with $p_A=1$.
Note that this choice of $p_A$ means that the mean binding time equals the expected time until two molecules occupy the same lattice for the first time and we are considering only diffusion-limited reaction events.

\begin{table}[h!]
\centering
\begin{tabular}{l|ll}
& Parameter & Mean binding time\\
\hline\\
BD & $k_A^{BD}=20.3$ & $\tau_A = 105.23\pm 1.06$ \\
CA & $p_A=1$ & $\tau_A = 103.81\pm 1.20$\\
RRE & $k_A = 3.8270$ & $\tau_A = 104.52$
\end{tabular}
\caption{The reaction rate $k_A^{BD}$ in Smoldyn and $k_A$ in the macroscopic RREs, such that the mean binding times agree with that of a CA simulation with $p_A=1$ in a dilute medium.}
\label{tab:BindingTime}
\end{table}

The corresponding macroscopic RRE for the concentration $a_d(t)$ of $A$ molecules in dilute medium is
\begin{equation}
\frac{d}{dt}a_{d}(t)=-k_Aa_d(t)b_d(t),
\end{equation}
where $k_A$ is the inverse of the mean binding time scaled by the system volume $V$.
If we assume that there are initially equally many $A$ and $B$ molecules, $a_{d0}=b_{d0}$, the solution is
\begin{equation}
a_d(t)=\frac{a_{d0}}{a_{d0}k_At+1}.
\label{eq:DiluteAss}
\end{equation}

\begin{figure}[h!]
\centering
\subfigure[$a_0=b_0=0.05$]{\includegraphics[width=.45\textwidth]{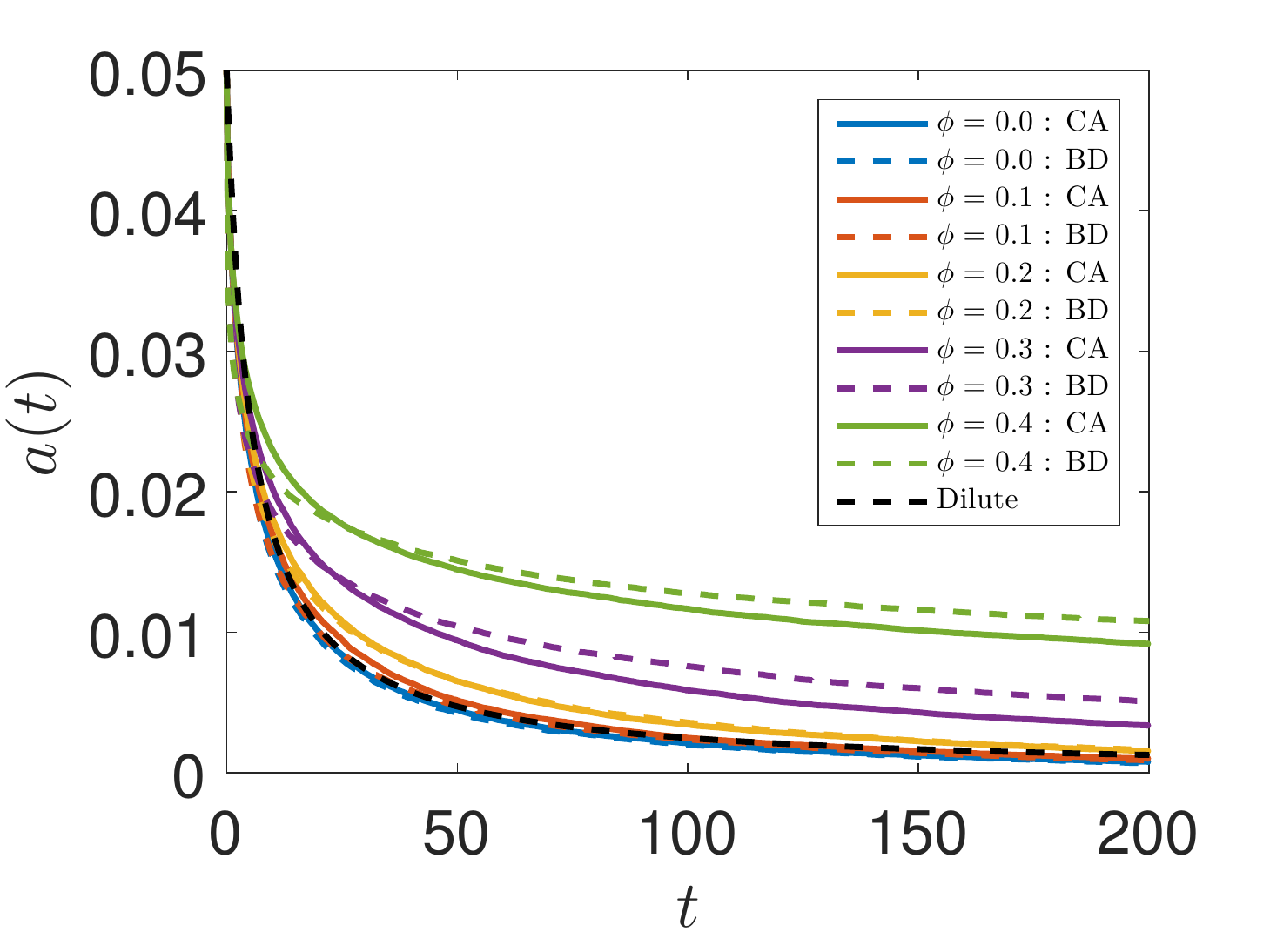}  }
\subfigure[$a_0=b_0=0.05$]{\includegraphics[width=.45\textwidth]{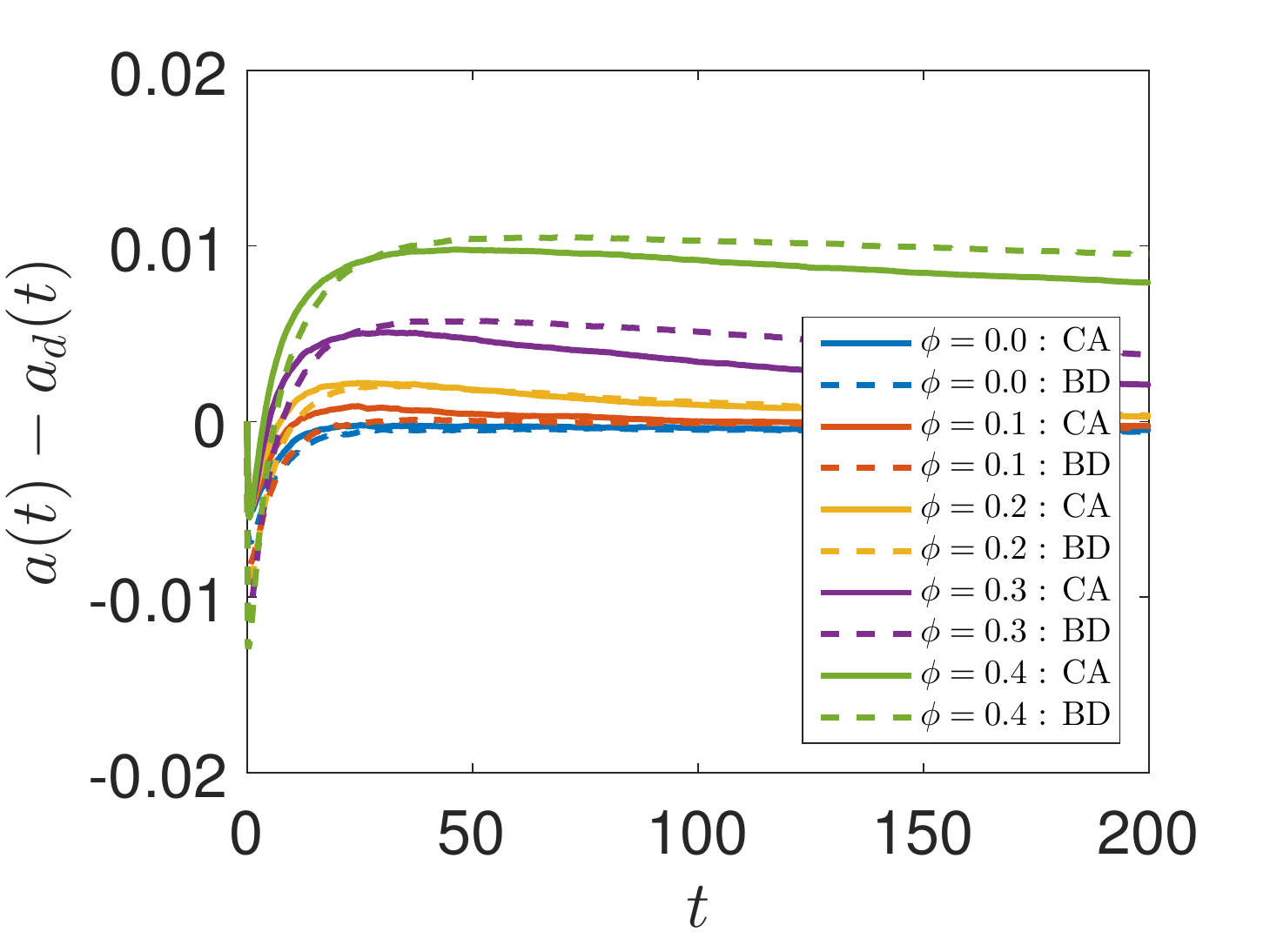}  }\\
\subfigure[$a_0=b_0=0.2$]{\includegraphics[width=.45\textwidth]{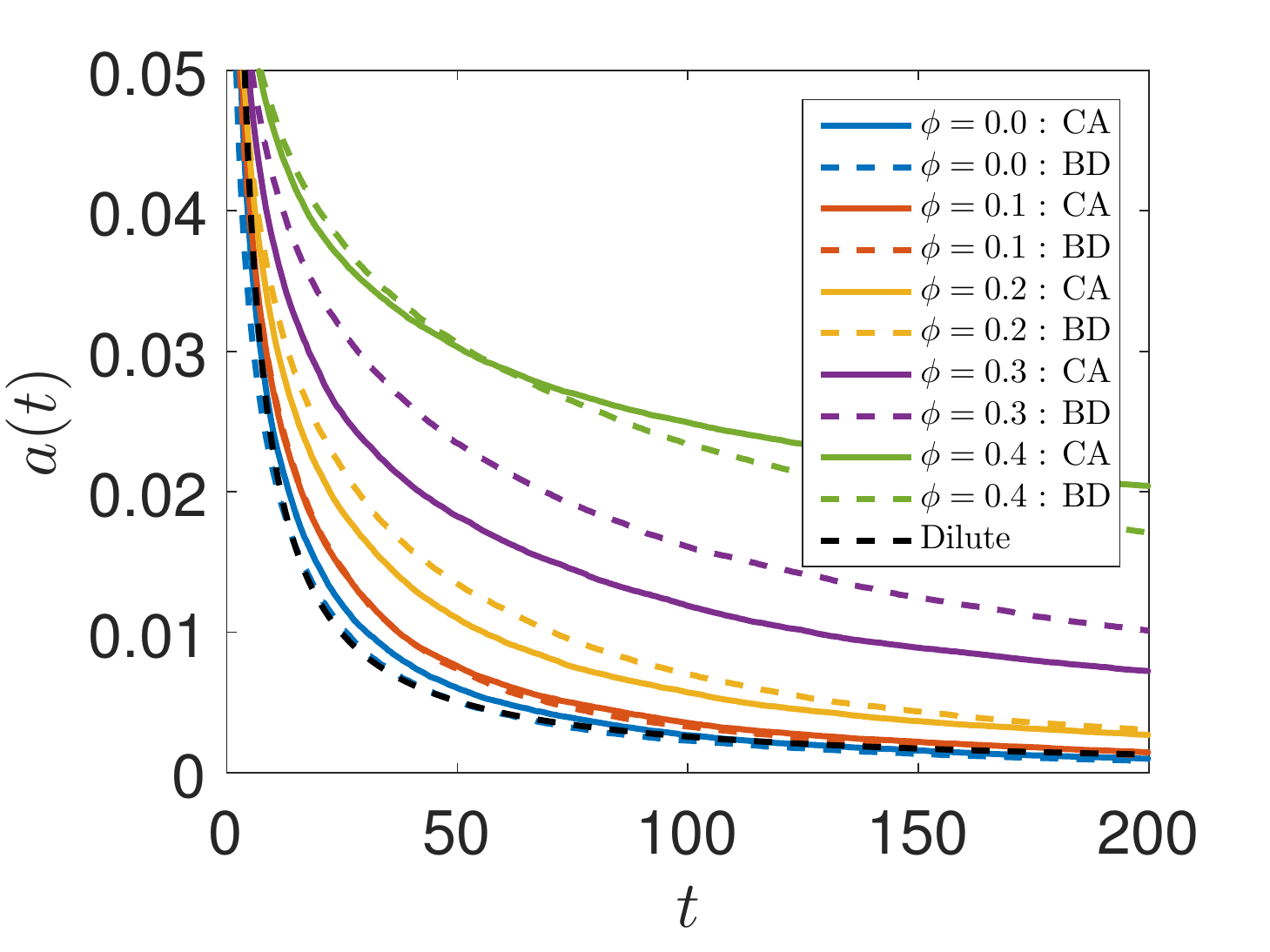}  }
\subfigure[$a_0=b_0=0.2$]{\includegraphics[width=.45\textwidth]{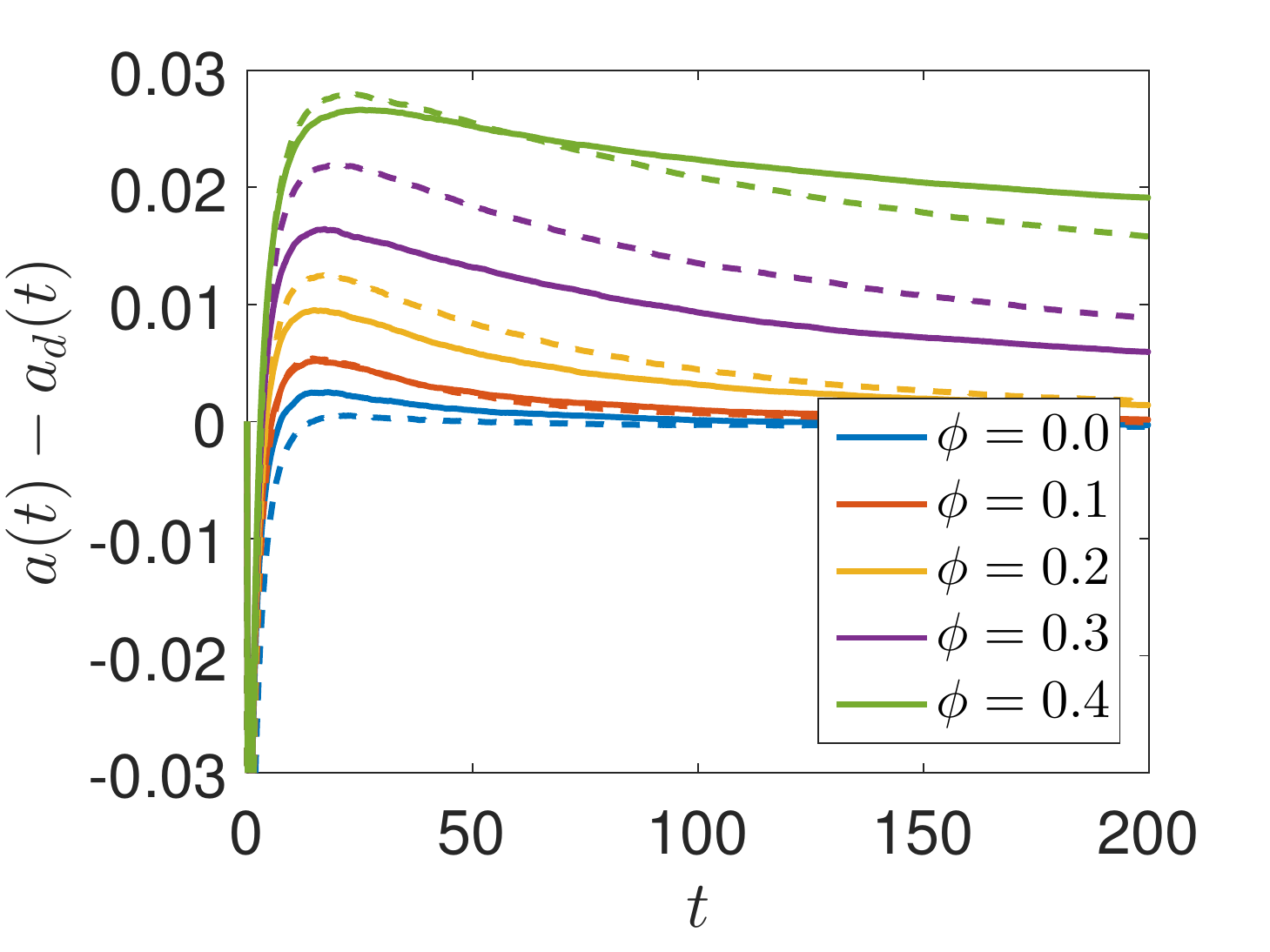}  }
\caption{Concentrations $a(t)$ of $A$ molecules for the association $A+B\to C$ in a crowded environment with volume occupancy $\phi$, simulated with CA (solid lines) and BD (dashed lines). First row: $a(t)$. Second row: $a(t)-a_d(t)$. For an increasing crowder density the reactions are slowed down compared to the dilute equation \eqref{eq:DiluteAss} and the difference between the on- and off-lattice models increases.}
\label{fig:Association}
\end{figure}

We introduce static crowders at different occupancies $\phi$ to the BD and CA simulations. 
Note that the produced $C$ molecules rest in the system as moving obstacles and here have the same size as $A$ and $B$ (Model I).
In Fig.~\ref{fig:Association} we plot the concentration $a(t)$ and the difference to the dilute solution $a_d(t)$ for each $\phi$.

\begin{figure}[h!]
\centering
\includegraphics[width=.7\textwidth]{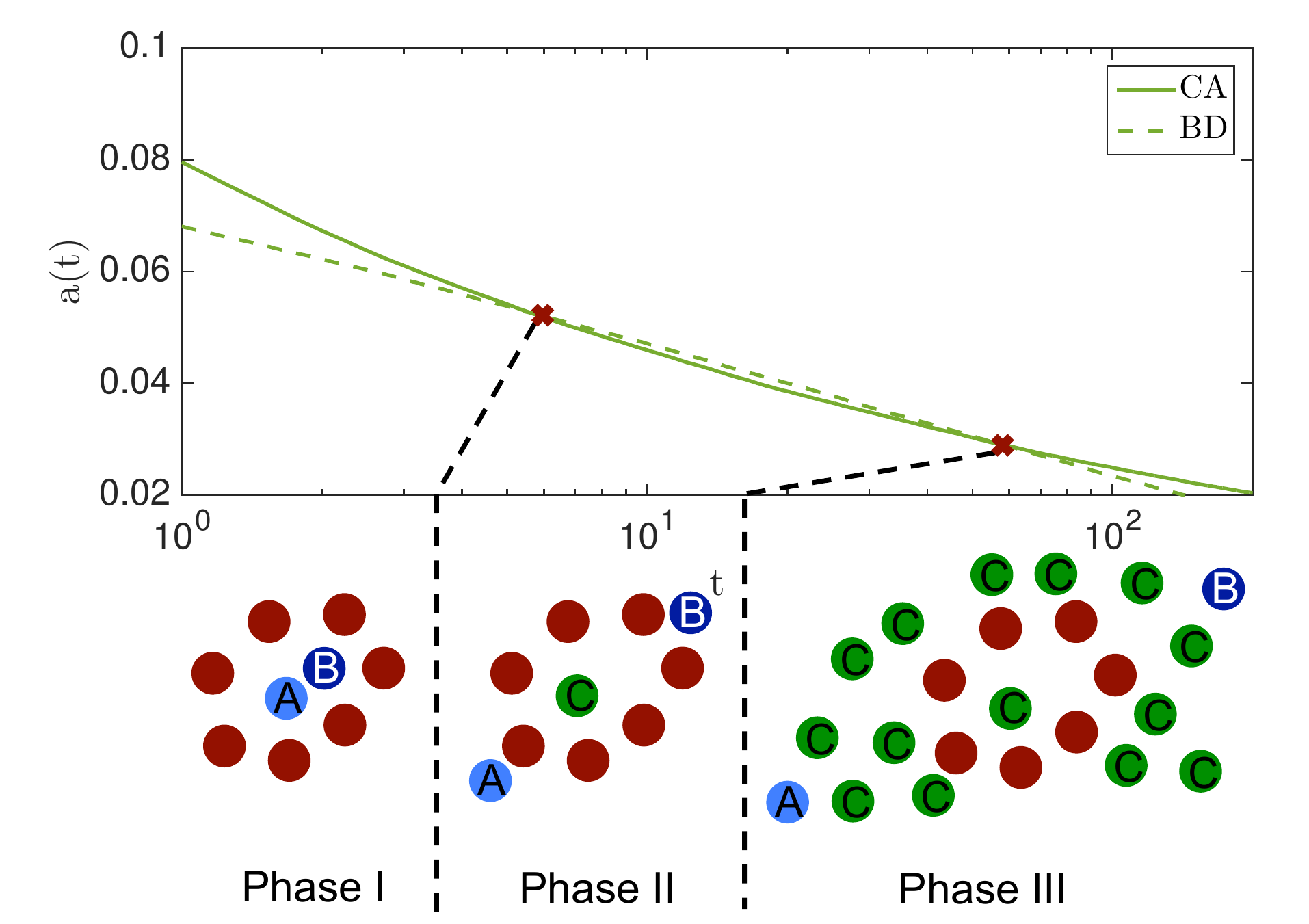} 
\caption{The concentration $a(t)$ of $A$ molecules plotted against $\log{t}$ simulated under high crowding concentrations with BD or CA. Phase I: Initially close pairs of $A$ and $B$ react faster in BD and hence $a(t)$ decreases faster with BD than with CA. Phase II: Distant pairs $A$ and $B$ encounter each other slower in BD and hence the BD reaction rate decreases compared to the CA reaction rate. Phase III: High density of moving obstacles $C$, which block the one-dimensional passageways forming in the CA geometry and hence separate $A$ and $B$ permanently from each other, while the moving $C$ allows $A$ and $B$ to collide more easily in the off-lattice simulations.}
\label{fig:CrowdingReactionsSchematic}
\end{figure}

For very short times the reaction speed is increased in Fig.~\ref{fig:Association}, because initially close reaction partners are kept in vicinity of each other by the surrounding obstacles.
But, generally the diffusion limited reactions simulated here are slower for higher crowder densities.
Introducing the artificial lattice, leading to a faster diffusion than in BD simulations, has a non-linear effect on the reaction rates.
For early times it slows down the reactions, since close reaction partners have a higher chance of escaping each other.
Then, a cross-over between the CA and BD curves occurs where the increased diffusivity of the on-lattice simulations leads to a faster encounter of initially distant reaction partners and the reaction rate is higher in the CA simulations.
The discrepancy between on- and off-lattice simulations increases for higher crowder densities and higher initial concentrations.
We call the latter effect self-crowding, meaning we can observe excluded volume effects even without adding explicit obstacles, since the $A$, $B$ and $C$ molecules themselves act as obstacles to one another.
Since $C$ is not actively participating in the reaction, we only simulate Model I, but the self-crowding effect of $C$ would be increased with Model II.
In Fig.~\ref{fig:CrowdingReactionsSchematic} we plot the concentration $a(t)$ simulated with both BD and CA in a highly crowded environment to explain the different phases. 
We only observe Phase III when the starting concentrations of $A$ and $B$ are $0.2$ and $\phi=0.4$, leading to the unbiological overall occupancy of $60\%$.
In this case the diffusivity in CA is as obstructed as in BD (see Fig.~\ref{fig:MSDstatic}(f)), but the grid makes it more difficult for molecules to pass each other, such that the moving $C$ molecules can permanently block $A$ and $B$ molecules from reacting, whereas in the off-lattice case there are more chances to pass moving obstacles.

\subsection{Dissociation events}
In this section we examine the dissociation reaction
\begin{equation}
C\xrightarrow{k_D} A+B
\label{eq:Diss}
\end{equation}
for the two different models for the size of $C$.
We first choose the dissociation probability in CA to be $p_D=0.1$ and then compute the mean time until a dissociation event happens by
\begin{equation}
\tau_D=\Delta t\sum_{k=1}^\infty p_D(1-p_D)^{k-1}k.
\end{equation}
From this we can derive the macroscopic reaction rate for the dissociation event
\begin{equation}
k_D=\frac{1}{\tau_D}=0.4
\end{equation}
and use it in the BD simulations.
In dilute medium the mean concentration of $A$ then follows the macroscopic RREs, which lead to the dilute concentration
\begin{equation}
a_d(t)=a_{d0}+c_{d0}\left(1-e^{-k_Dt}\right).
\label{eq:DissRRE}
\end{equation}

\begin{figure}[t!]
\centering
\subfigure[$c_0=0.1$]{\includegraphics[width=.45\textwidth]{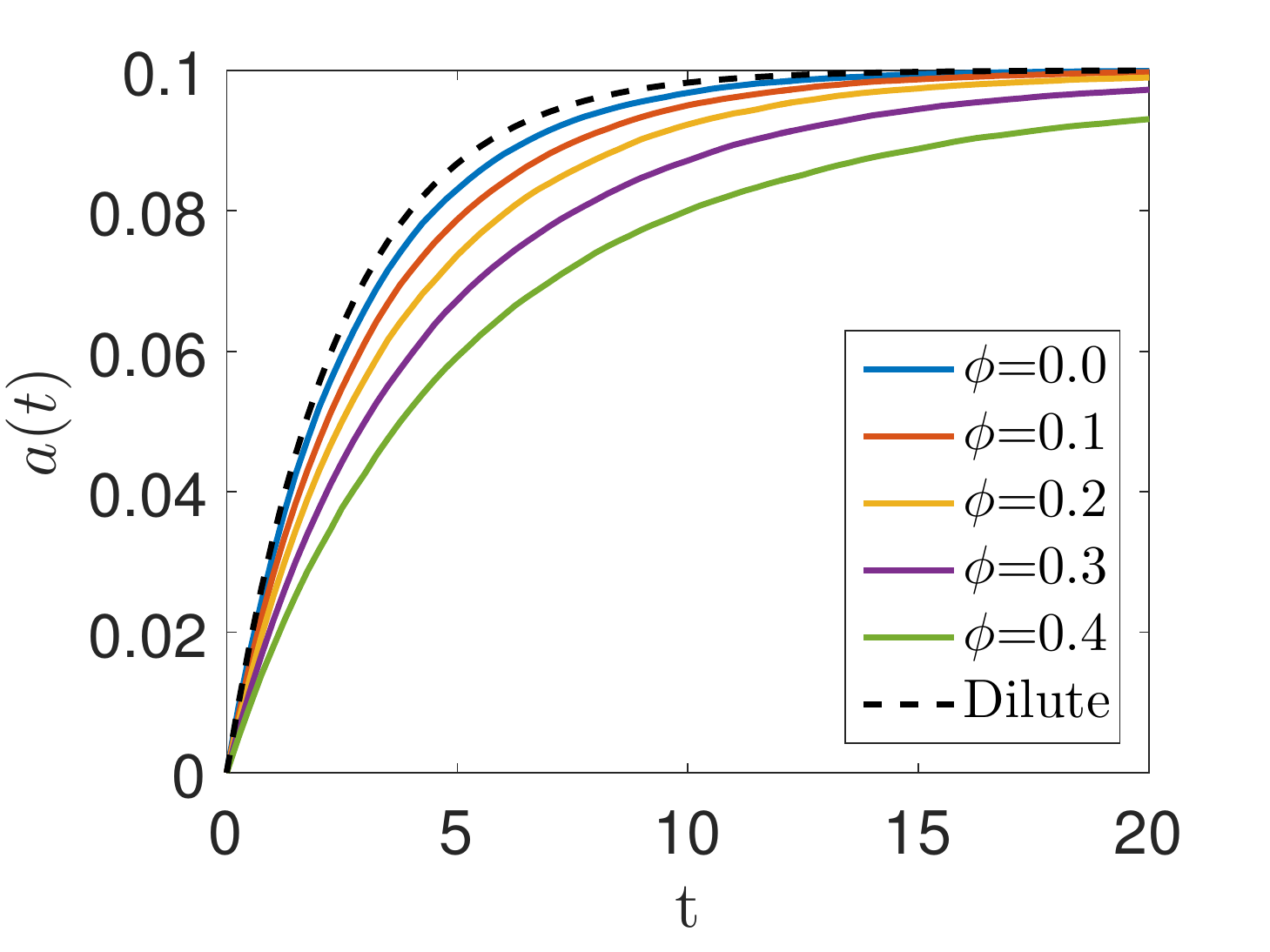} }
\subfigure[$c_0=0.1$]{\includegraphics[width=.45\textwidth]{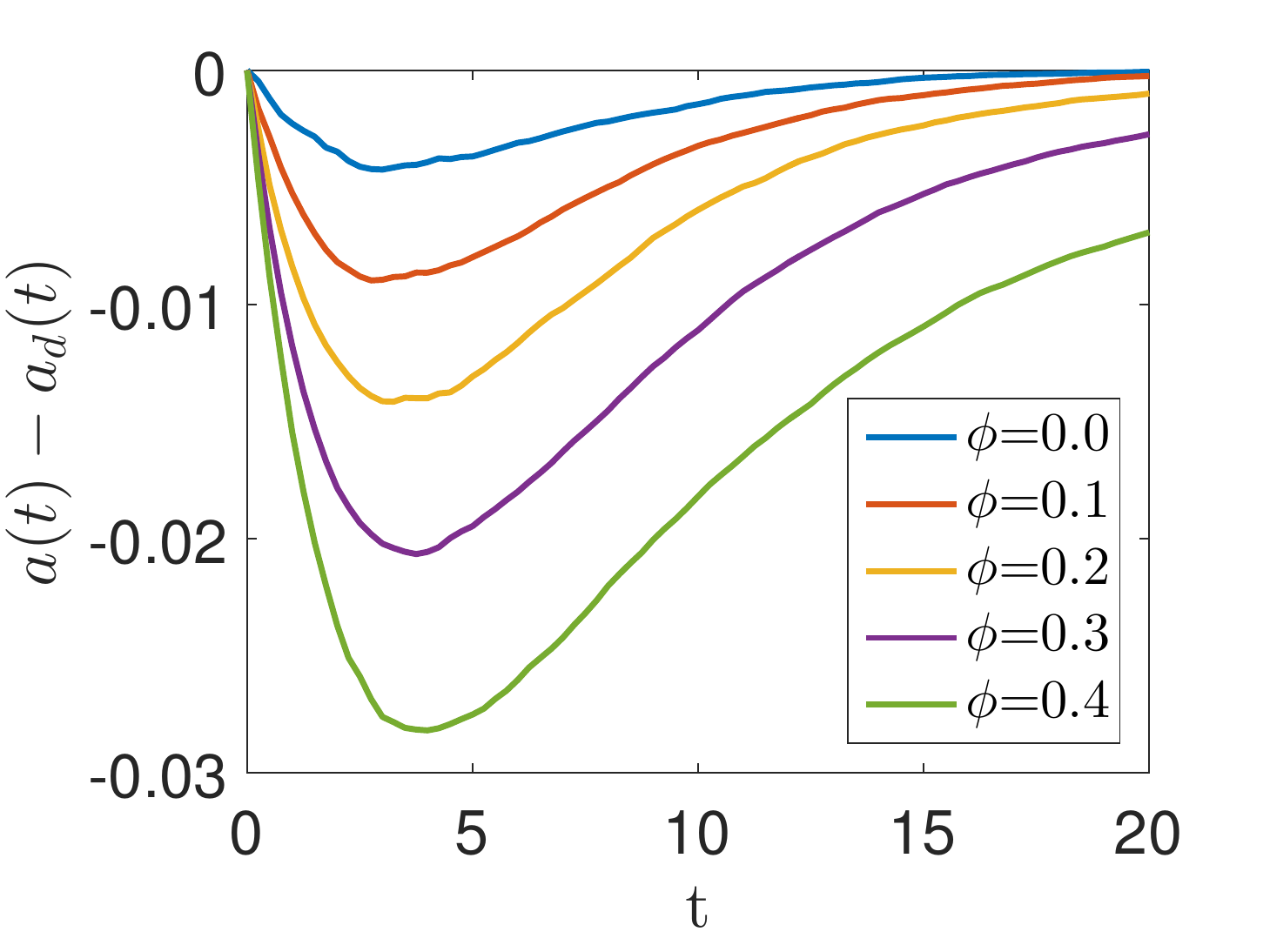} }\\
\subfigure[$c_0=0.4$]{\includegraphics[width=.45\textwidth]{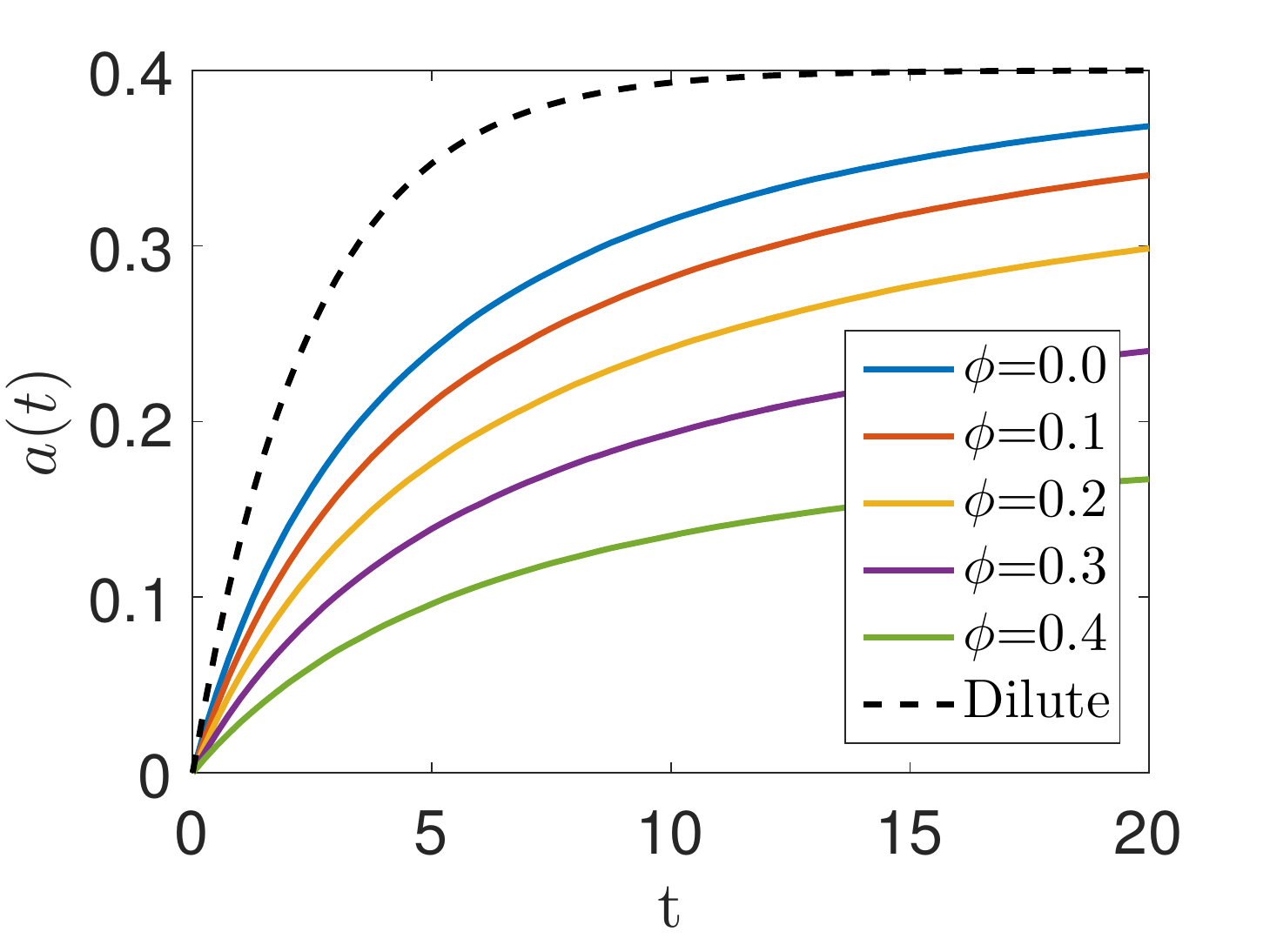} }
\subfigure[$c_0=0.4$]{\includegraphics[width=.45\textwidth]{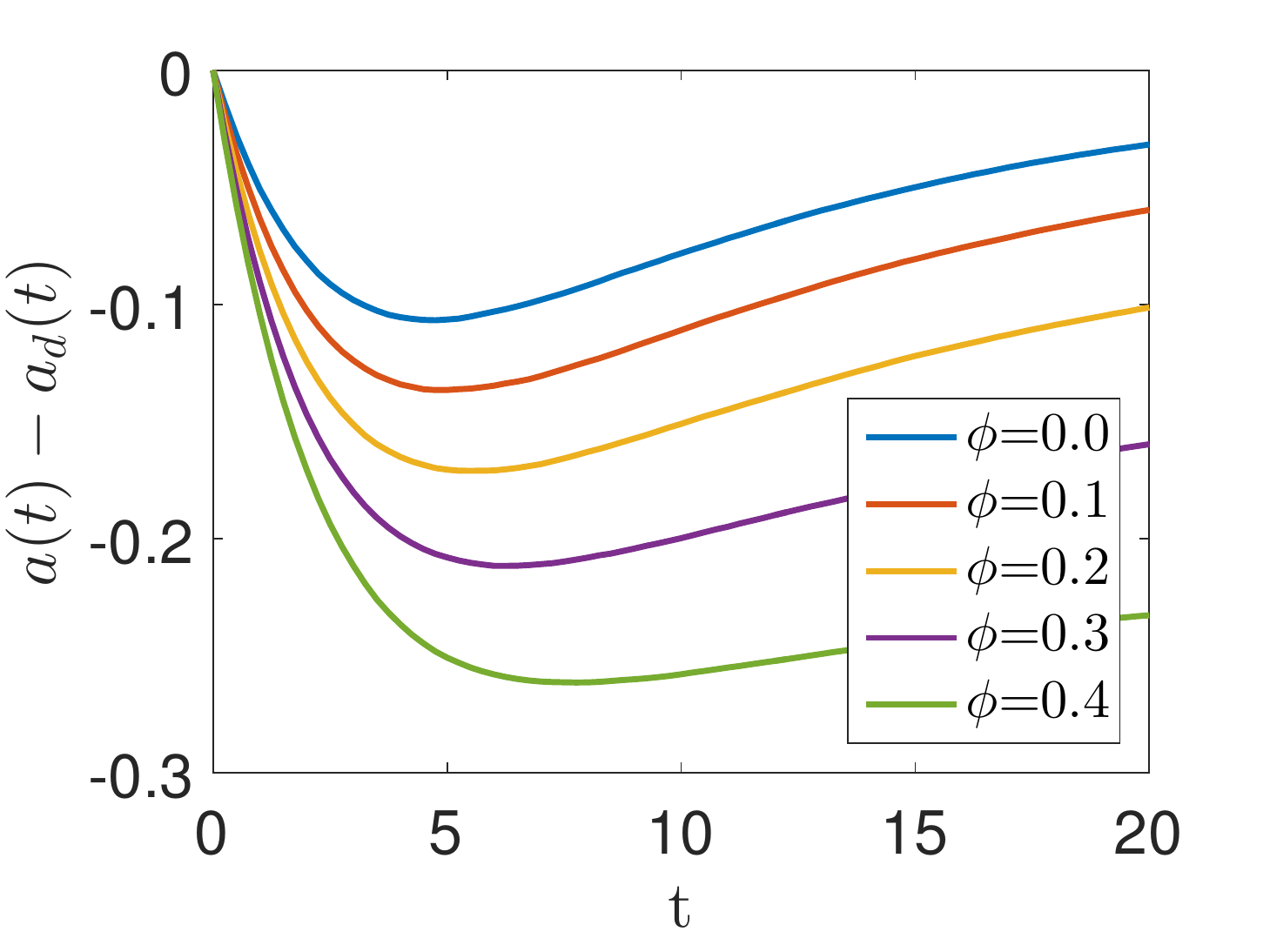} }
\caption{Concentration $a(t)$ of the $A$ molecules in a system with the dissociation reaction $C\to A+B$ and Model I simulated with CA. We observe a self-crowding effect, since the simulations with $\phi=0$ differ from the theory for dilute media and this effect increases with increasing initial concentration. Since an additional lattice site is needed for the dissociation to be successful, excluded volume effects considerably obstruct this reaction, when modeled with Model I, and $a(t)$ decreases for increasing $\phi$.}
\label{fig:Diss1}
\end{figure}

When $C$ follows Model I extra space is needed to place one of the reaction products $A$ or $B$ and the dissociation might be rejected if another molecule occupies the position sampled for the extra particle.
This effect, however, is not possible to achieve with the software Smoldyn, as it always places the new particle in the system and decides in the next time step if a possible rebinding or a diffusive jump happen.
Hence, we only examine CA simulations for different fractions of occupied volume $\phi$ and compare them to the analytic solution in the dilute medium in Fig.~\ref{fig:Diss1}.

Again, we observe self-crowding effects, as the simulations in the dilute medium for $\phi=0$ do not follow the dilute concentration \eqref{eq:DissRRE}.
The $A$ and $B$ molecules here act as moving obstacles that that block the extra lattice site needed for the dissociation of $C$, and we see that the discrepancy between the simulations and the analytic solution \eqref{eq:DissRRE} is more pronounced for an increased initial concentration $c_0$.
Introducing obstacles further stabilizes the complex $C$, so that fewer $A$ and $B$ molecules are created, leading to a lower concentration $a(t)$.

\begin{figure}[h!]
\centering
\includegraphics[width=.45\textwidth]{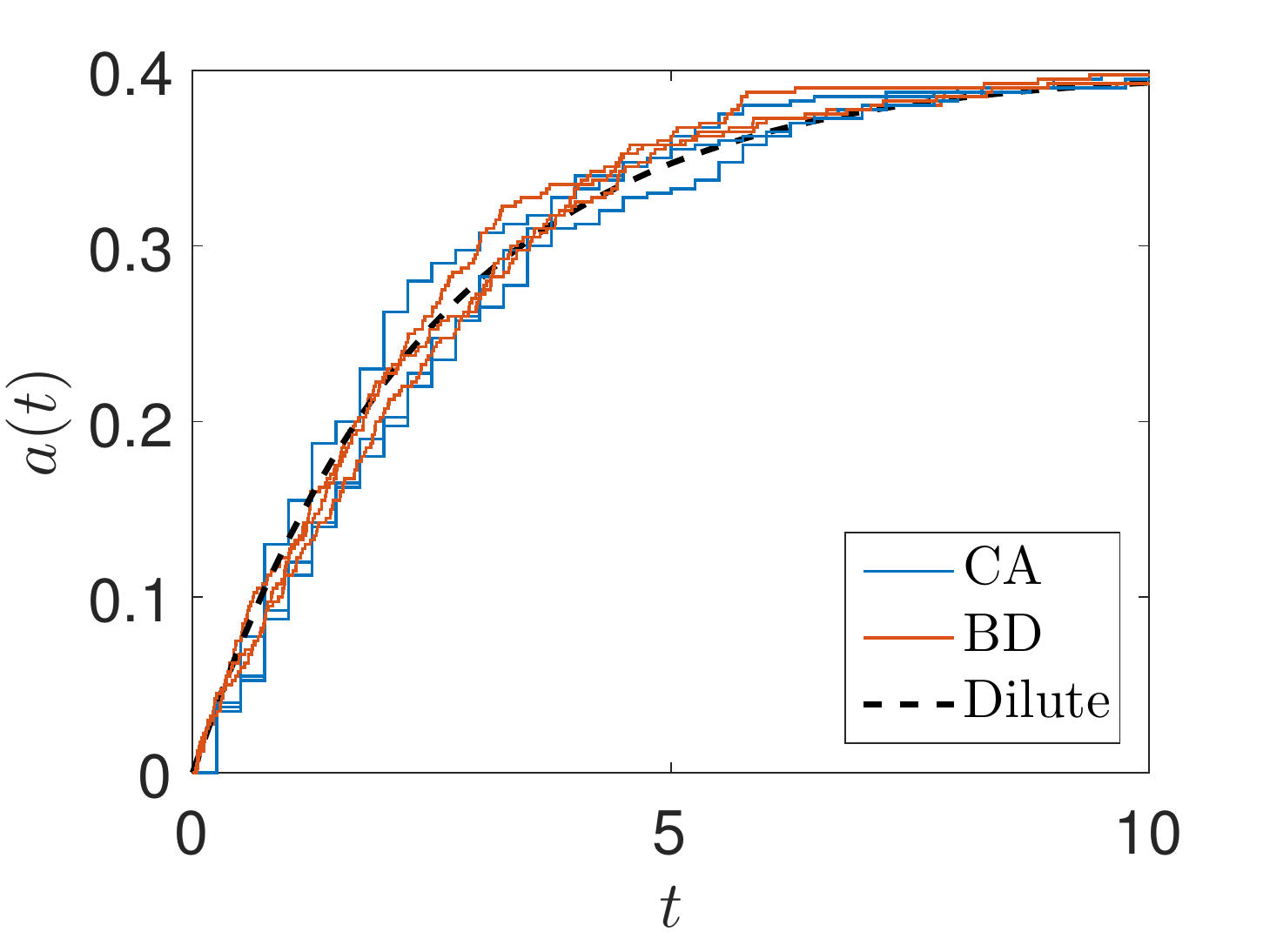} 
\caption{Sample paths of the dissociation reaction $C\to A+B$, where $C$ follows Model II, simulated with BD and CA, and the macroscopic solution \eqref{eq:DissRRE} in dilute medium. For Model II no extra space is needed to place the newly created molecule and hence on- and off-lattice models follow the theory for dilute media for all $\phi$. This differs considerably from the simulation results with Model I, see Fig.~\ref{fig:Diss1}}
\label{fig:Diss2}
\end{figure}

When $C$ is double the size of $A$ and $B$ on the other hand (Model II), the dissociation event is always successful and independent of the occupancy $\phi$, since no additional site is needed.
In Fig.~\ref{fig:Diss2} we show some sample paths of both BD and CA simulations, they both agree with the analytic concentration in \eqref{eq:DissRRE}.
Thus, the choice of model for $C$ plays an important role for the outcome of the dissociation reaction and macromolecular crowding has no effect on the outcome of a dissociation reaction in discrete and continuous space models if the more realistic Model II is chosen.
This model is natural to implement in BD, but leads to the unnatural representation of $C$ in CA, where molecules generally are assumed to all have the same size.

\subsection{Reversible reactions}
We now combine our results to examine the reversible binding reaction
\begin{equation}
A+B\xrightleftharpoons[k_D]{k_A} C,
\label{eq:Reverse}
\end{equation}
where we choose the reaction constants in the same way as in the previous sections.
To guarantee the correct rebinding probability in the Smoldyn simulations, we manually set it to $0.25$ to agree with the rebinding probability in CA, where a newly produced particle jumps with probability $0.25$ into the adjacent reaction partner and they react with $p_A=1$.
The RREs for \eqref{eq:Reverse} are non-linear and hence not analytically solvable, but we can derive expressions for the steady state concentration $\bar{a}$ of $A$
\begin{equation}
\bar{a}=\frac{1}{2}\left(-\frac{k_D}{k_A}+\sqrt{\left(\frac{k_D}{k_A}\right)^2+4\frac{k_D}{k_A}(a_0+c_0)}\right).
\label{eq:SteadyState}
\end{equation}

\begin{figure}[h!]
\centering
\subfigure[$a_0=0.05$]{\includegraphics[width=.45\textwidth]{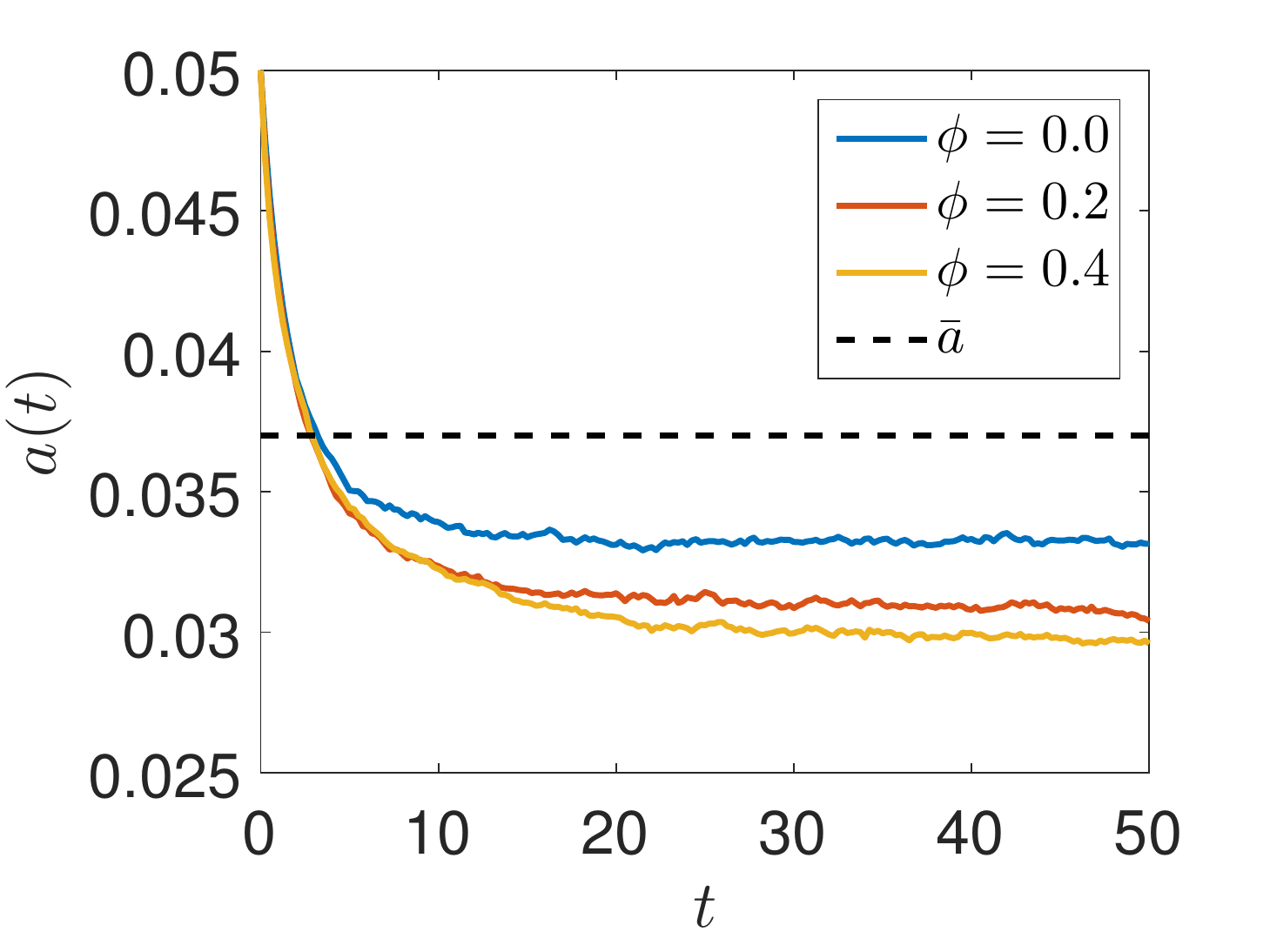} }
\subfigure[$a_0=0.2$]{\includegraphics[width=.45\textwidth]{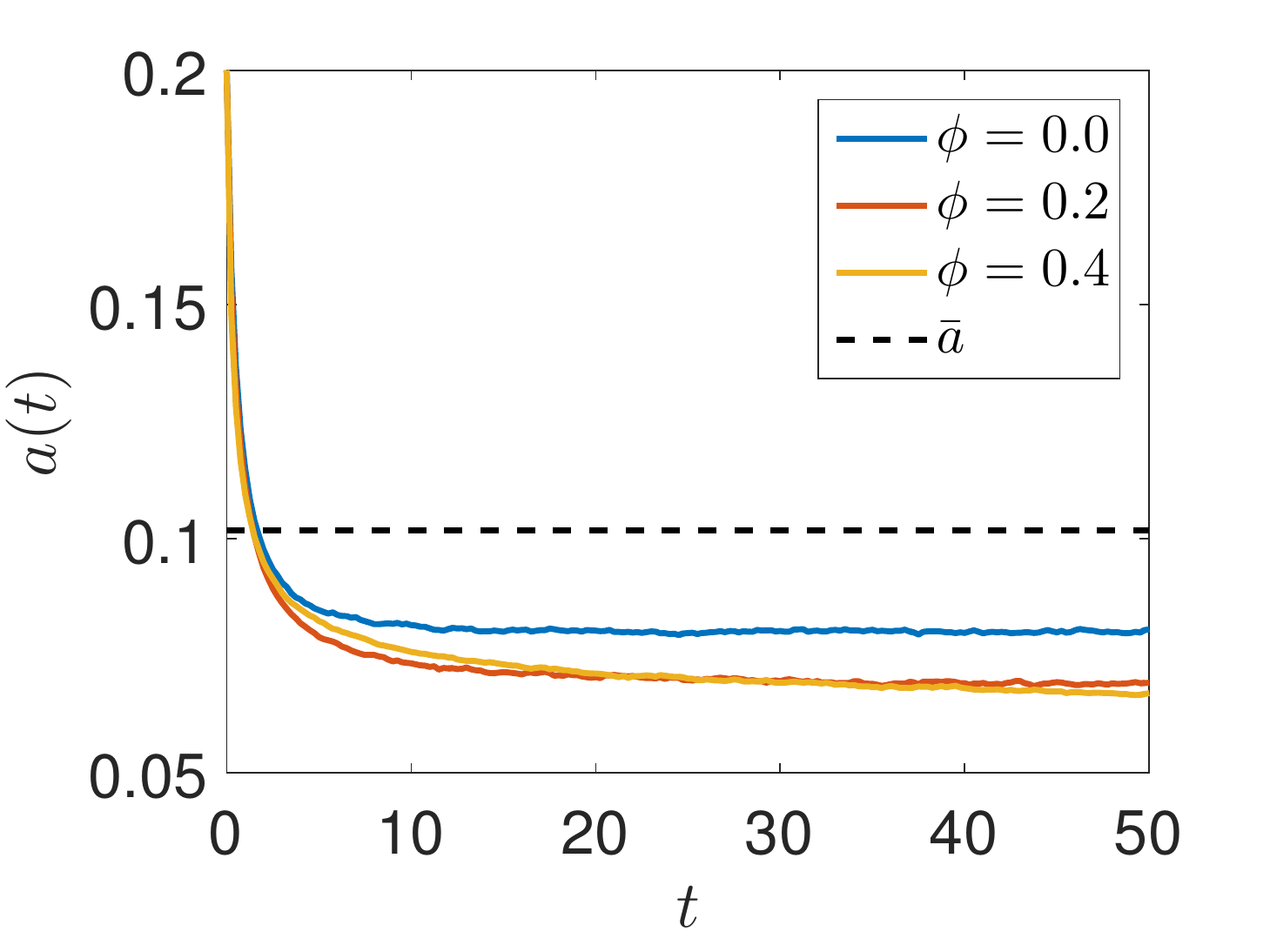} }
\caption{The concentration of $A$ molecules for the reversible reaction \eqref{eq:Reverse} simulated with CA, for Model I and the steady state solution $\bar{a}$ in dilute medium \eqref{eq:SteadyState}. The self-crowding effect for $\phi=0$ stabilizes the complex $C$, leading to a lower concentration of $A$ molecules than predicted by the theory for dilute media. This effect is increased when additional crowding molecules are introduced into the system.}
\label{fig:Reverse1}
\end{figure}

We first investigate Model I for $C$ by simulating the reversible reaction only with CA for different $a_0=b_0$ and $c_0=0$.
In Fig.~\ref{fig:Reverse1} we observe that increasing crowding stabilizes the complex $C$ and decreases the amount of $A$ and $B$ molecules in the system at steady state as compared to the dilute case \eqref{eq:SteadyState}.
There are two reasons for this: (i) a particle already occupies the lattice for the newly created molecule and the dissociation event is rejected; (ii) the rebinding time for $A$ and $B$ is decreased, since they escape from each other more rarely.
The decreased time for $A$ and $B$ to meet in a crowded environment does here not effect the steady state concentrations.
The self crowding effect is of the same order as the crowding effect, where the simulated steady state for $\phi=0$ is $\sim 0.9\bar{a}$ for $a_0=0.05$ and $\sim0.79\bar{a}$ for $a_0=0.2$,  indicating that the $A$ and $B$ molecules themselves heavily stabilize the complex $C$.

\begin{figure}[h!]
\centering
\subfigure[$a_0=0.05$]{\includegraphics[width=.45\textwidth]{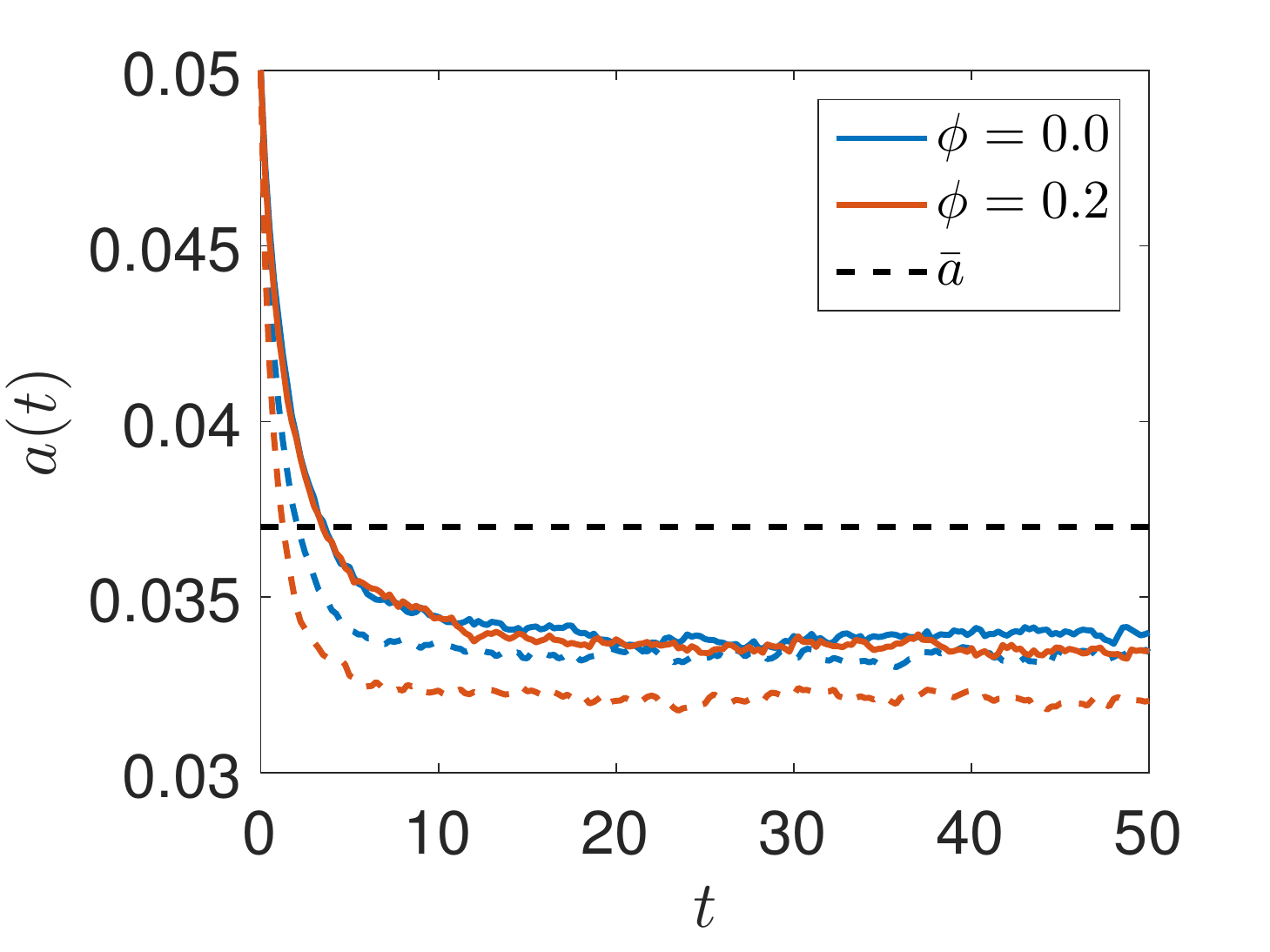} }
\subfigure[$a_0=0.2$]{\includegraphics[width=.45\textwidth]{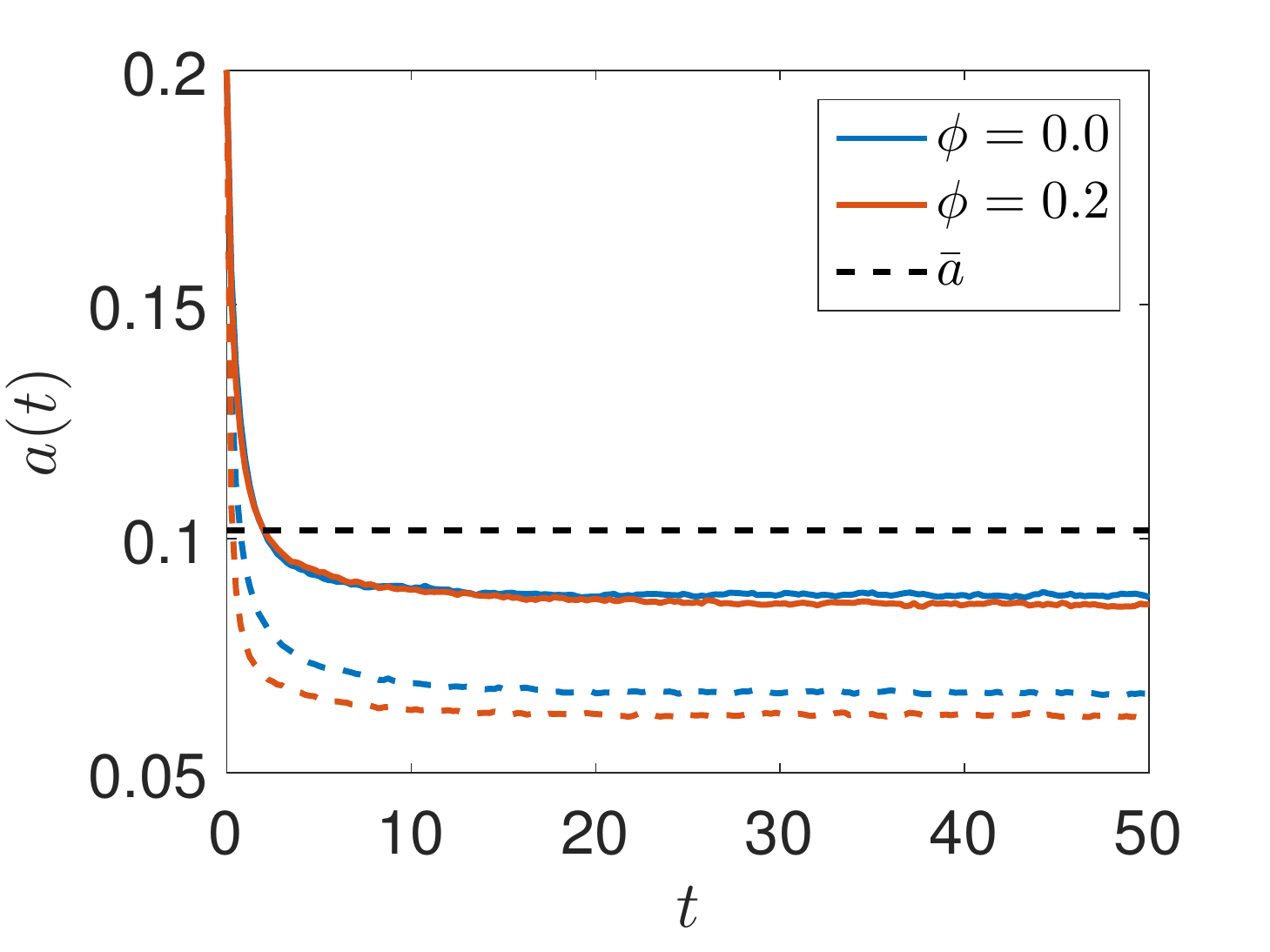} }
\caption{The concentration $a(t)$ of $A$ molecules for the reversible reaction \eqref{eq:Reverse} with Model II, simulated with CA (solid lines) and BD (dashed lines). No additional space is needed to place the new molecule when the complex $C$ decays, but since the steady state concentration of $A$ is lower than the one predicted for dilute media, we can conclude that it is mostly the decreased rebinding time for $A$ and $B$, that leads to the difference between the theory for dilute media and the simulated steady states in a crowded medium. This effect is stronger when simulated off-lattice, since it is more difficult for newly created $A$ and $B$ molecules to escape from each other.}
\label{fig:Reverse2}
\end{figure}

To investigate the grid effect on the reversible reaction, we will now perform the same experiments for Model II, Fig.~\ref{fig:Reverse2}.
As compared to the CA simulations for Model I we see that the steady state levels are slightly elevated, which is expected since the complex $C$ already occupies the space needed to dissociate into $A$ and $B$.
The steady state level, however, still lies beneath the analytically predicted one for dilute media, indicating that the predominant effect is the decreased rebinding time of $A$ and $B$ in a crowded environment.
The on-lattice simulations here underestimate the excluded volume effect as it is easier for $A$ and $B$ to diffuse away from each other.
In the case of the reversible reaction, investigated here, the grid artifact on the reaction rates is linear with no cross-over between CA and BD, and increases with increasing $\phi$.
The computational time between the two models differs considerably, for the case with $a_0=b_0=0.2$ the CA simulations were ca. 42 faster than the more accurate BD simulations.

\begin{figure}[h!]
\centering
\subfigure[BD: $\phi=0.0$]{\includegraphics[width=.285\textwidth]{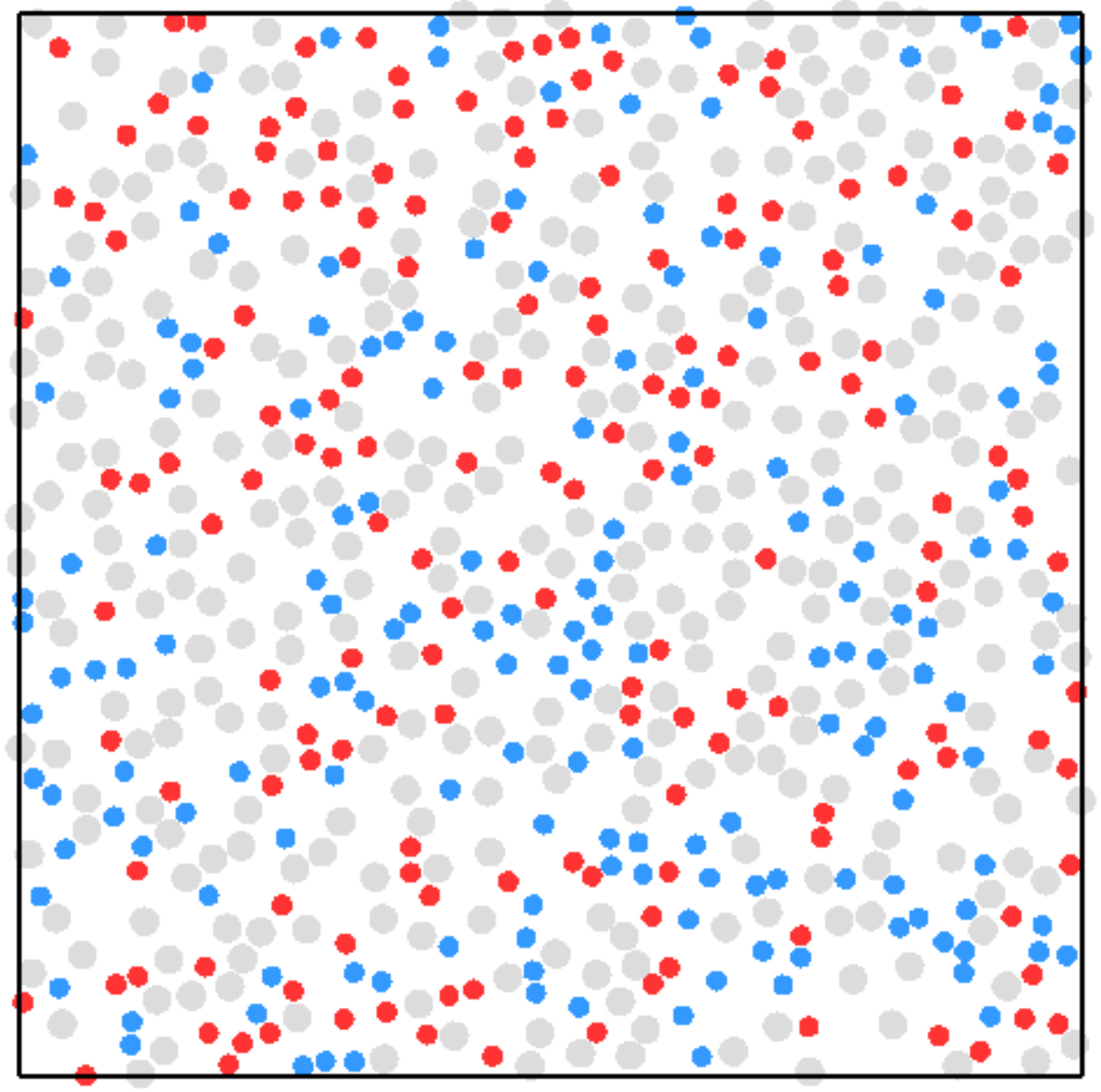} }\hspace{.8cm}
\subfigure[BD: $\phi=0.2$]{\includegraphics[width=.285\textwidth]{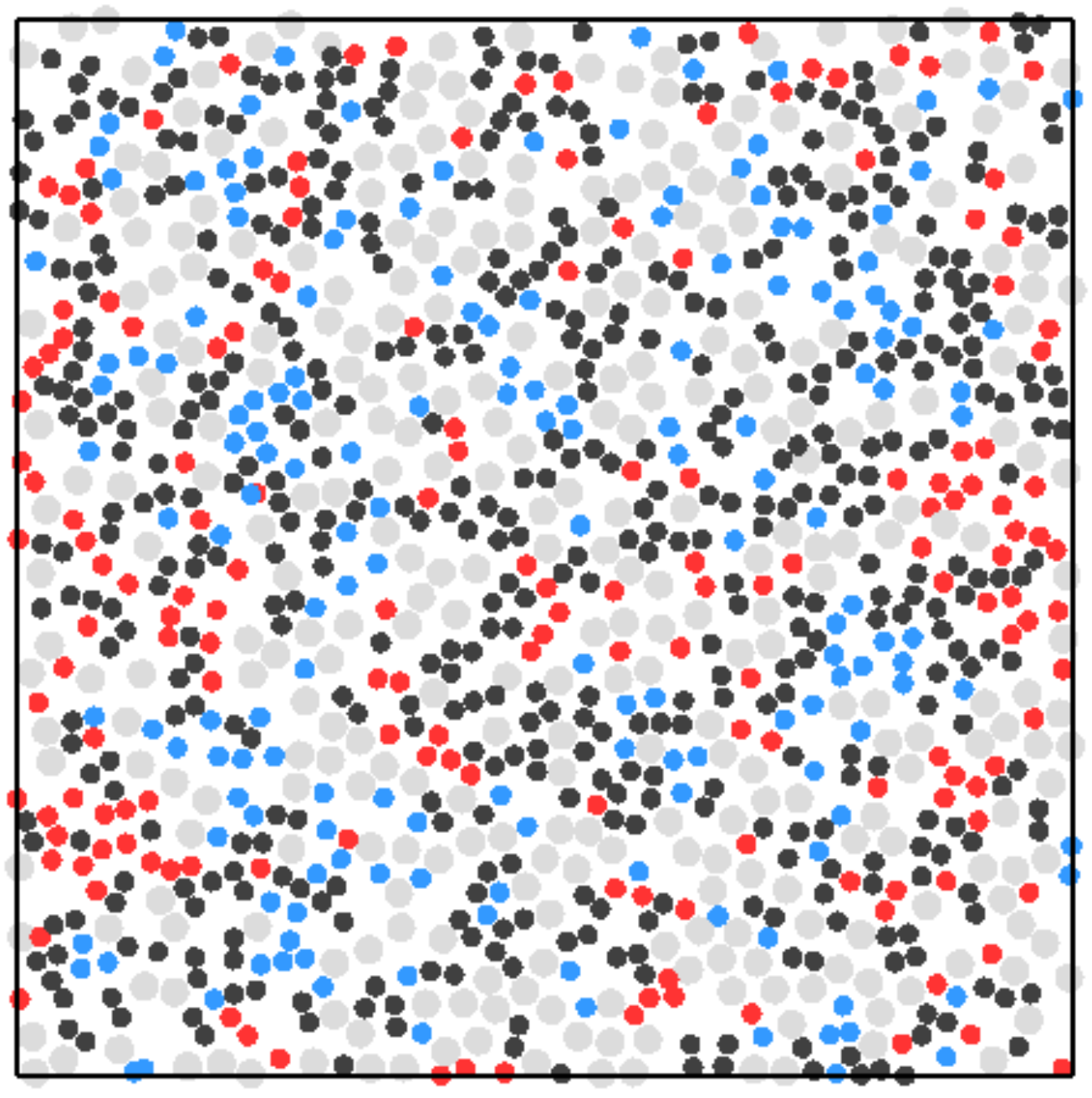} }\\
\subfigure[CA: $\phi=0.0$]{\includegraphics[width=.35\textwidth]{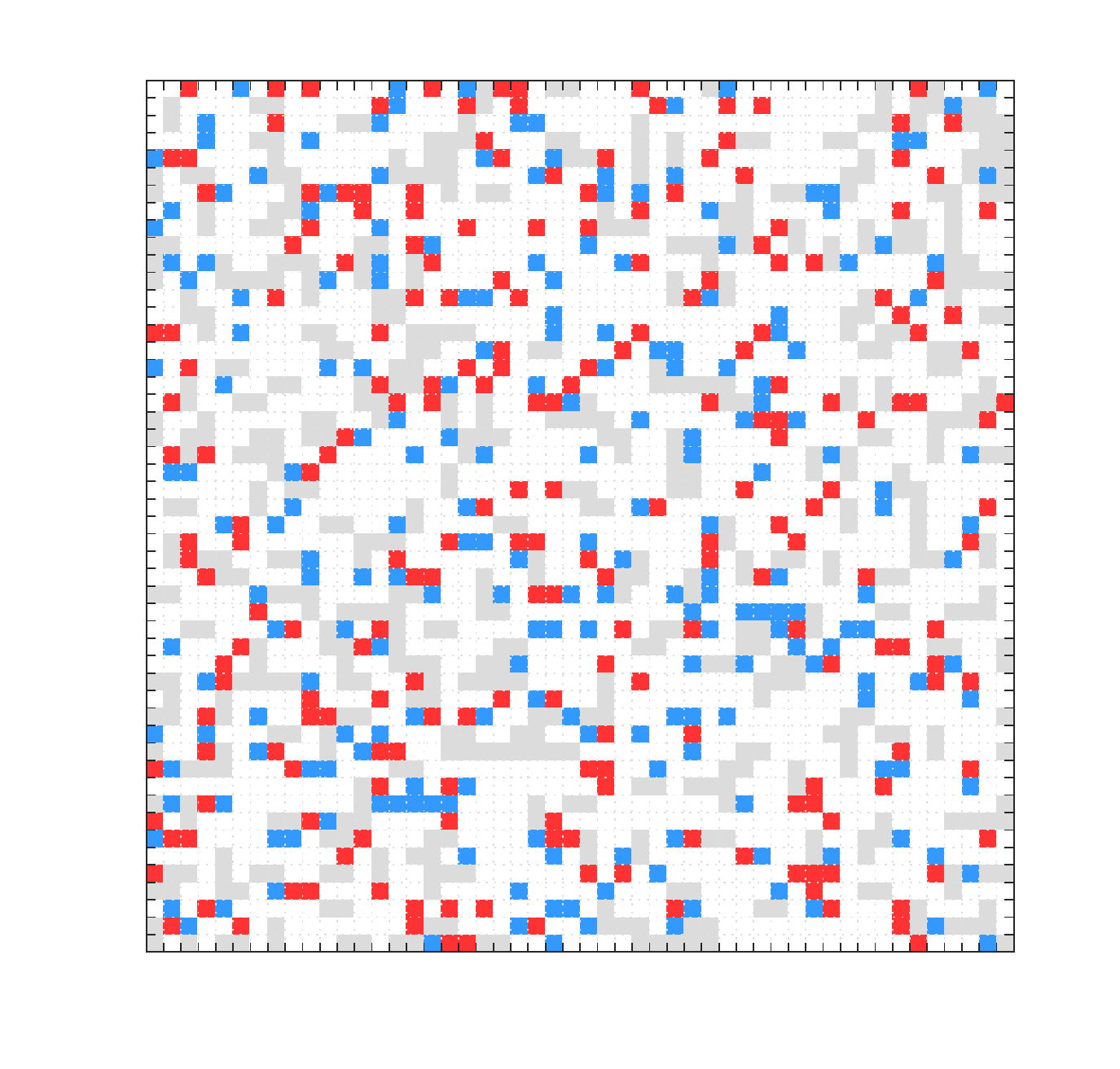} }
\subfigure[CA: $\phi=0.2$]{\includegraphics[width=.35\textwidth]{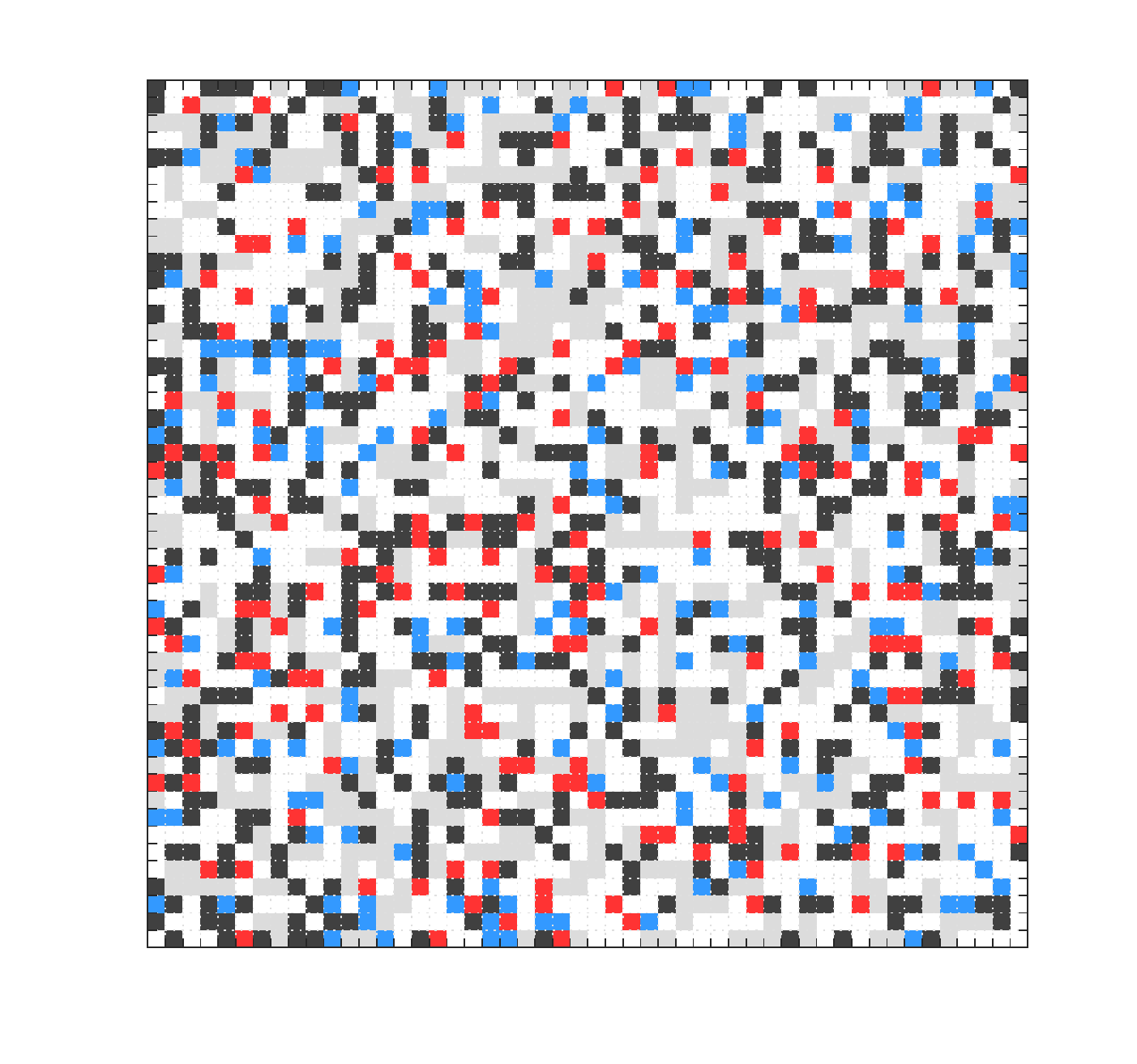} }
\caption{The spatial distributions of $A$ (blue), $B$ (red) and $C$ (grey) and the static obstacles (black) in the steady state of the reversible reaction \eqref{eq:Reverse} and initial concentrations $a_0=b_0=0.2$ and $c_0=0$. With increased crowder density $\phi$ the spatial heterogeneity increases.}
\label{fig:Distribution}
\end{figure}

Last, we investigate the variance in the reversible reaction system and how it depends on the level of crowding.
We performed the same experiments as for Fig.~\ref{fig:Reverse2}, but with only 100 trajectories in one crowder distribution, but there was no visible difference in the variance of the overall concentration of $A$.
In Fig.~\ref{fig:Distribution} we depict snapshots of the distributions of $A$, $B$ and $C$ molecules together with the obstacles when the system has reached steady state.
Here, we observe that an increased crowder density leads to spatially more inhomogeneous distributions of $A$ and $B$, since close reactants are stabilized in the complex $C$. 

\newpage
\section{Conclusions}\label{sec:Conclusions}
To understand complex gene regulatory networks in the crowded cell it is essential to perform realistic and computationally efficient reaction-diffusion simulations capturing the excluded volume effects, due to the high concentration of crowding macromolecules.
In this paper we perform rigorous reaction-diffusion simulations in discrete and continuous space to compare how well the on-lattice models approximate the more accurate off-lattice models, when applied to crowded environments.
Due to the computational complexity we hereby restrict our study to static crowders, to diffusion-limited reactions with reaction probability one, and to the two-dimensional case, providing insight into reaction-diffusion processes on biological membranes. 

In a pure diffusion system we first observe that all models result in slower diffusion for an increased fraction of occupied volume, as was expected.
This effect, however, is more pronounced with off-lattice simulations, which was not expected as they are more flexible in the number of directions a molecule can move.
A possible explanation to this phenomenon is, that the artificial grid orders the particles such that they effectively excluded less space.
The consequences of the lattice on the outcome of diffusion limited association reactions is twofold: for short times the bimolecular reaction rates are decreased by the artificial grid, but for long times they are increased due to the faster hitting times between the reactants.
For a reversible reaction we find that the on-lattice simulations underestimate the crowding effects compared to the more detailed BD model.
In all experiments the excluded volume effects increase for higher initial concentrations of the reactants, indicating that they themselves act as crowders, an effect we call self-crowding.
When modeling dissociation or reversible reactions we illustrated that it is important to model the molecules with their actual size,  a feature usually not considered in the CA models, where all molecules are assumed to be the size of one lattice site.
On the other hand, we have given examples of the considerable decrease in computation time for the CA model as compared to the costly off-lattice simulations.

We observe that the overall variance of the concentration of molecules is independent of the presence of crowding molecules, but the spatial variation in a reversible reactive system is increased, since close reactants are more likely to be bound in a complex and distant reactants are more hindered to meet each other.

The main aim of this article is to investigate the accuracy of on-lattice approximations to BD simulations, since they are popular for investigating excluded volume effects.
We find that the artificial lattice in the computationally more efficient CA model has significant artifacts especially on the diffusive behavior and the steady-state concentration of a reversible reaction.
Hence it can be regarded as an alternative to the more accurate off-lattice simulations only for low crowder distributions.

\section*{Acknowledgements}
This work was supported by the Swedish Research Council grant 621-2001-3148.
The authors would like to thank Per L\"{o}tstedt for his support and Andreas Hellander for helpful comments on this manuscript.

\bibliographystyle{plain}
    \bibliography{micro}

\end{document}